\documentclass[pra,reprint,longbibliography]{revtex4-1}

\usepackage{graphicx}
\usepackage{physics}
\usepackage{mathrsfs}

\usepackage{mathtools}
\newcommand{\e}{\mathrm{e}}
\newcommand{\Cramer}{Cram\'er }
\newcommand{\RLD}{L^{(\text{R})}}
\newcommand{\SLD}{L^{(\text{S})}}
\usepackage{amsmath,amsfonts,amssymb}

\usepackage{hyperref}

\hypersetup{
 linktocpage=true,
 pdfborderstyle={/S/S/W 1},
 hyperindex=true,
 bookmarks=true,
 bookmarksopen=true,
 bookmarksnumbered=true,
}

\newtheorem{program}{Program}
\begin{document}

\title{Ultimate precision of joint quadrature parameter estimation with a Gaussian probe}
\author{Mark Bradshaw, Ping Koy Lam, Syed M. Assad}
\affiliation{Centre for Quantum Computation and Communication Technology, Department of Quantum Science,\\ Research School of Physics and Engineering, Australian National University, Canberra ACT 2601, Australia.}

\begin{abstract}
The Holevo Cram\'er-Rao bound is a lower bound on the sum of the mean-square error of estimates for parameters of a state.
  We provide a method for calculating the Holevo Cram\'er-Rao bound
  for estimation of quadrature mean parameters of a Gaussian state by
  formulating the problem as a semidefinite program. In this case, the bound is tight; it is attained by purely Guassian measurements. We consider the example of a symmetric two-mode squeezed thermal state undergoing an unknown displacement on one mode. We calculate the Holevo Cram\'er-Rao bound for joint estimation of the conjugate parameters for this displacement. The optimal measurement is different depending on whether the state is entangled or separable.
\end{abstract}

\maketitle
\section{Introduction}

Quantum mechanics sets a limit on how accurately one can measure two
noncommuting observables. This is exemplified by the Heisenberg
uncertainty relation for position and momentum, which can be
generalized to arbitrary observables~\cite{PhysRev.34.163}. This
relation sets a precision limit to state estimation problem of the noncommuting
observables. For example if we were to simultaneously measure two
quadrature operators $\mathcal{Q}$ and $\mathcal{P}$ with the
canonical commutation relation
$[\mathcal Q,\mathcal
P]=i$~\cite{weedbrook2012gaussian,adesso2014continuous} of a quantum
state $\rho$, then the precision is limited by
$\Delta{\mathcal{Q}}\Delta{\mathcal{P}} \geq \frac{1}{2}$. However if we
are interested in estimating channel parameters instead, this restriction
do not apply. In this case, entanglement can be used to enhance the precision of channel parameter estimates~\cite{Giovannetti1330,PhysRevLett.87.270404,fujiwara2001quantum,fischer2001enhanced,sasaki2002optimal,fujiwara2003quantum,ballester2004estimation}, for example, estimating the squeezing applied to a probe~\cite{rigovacca2017versatile}. Light-matter interferometry can be used to improve the estimate of a Gaussian process applied to a matter system~\cite{ruppert2017light}. The precision can also be improved with a cleverly chosen single-mode state, for the estimation of a small displacement, for example~\cite{duivenvoorden2017single}. 

We will consider in detail the example of estimation of the parameters $\theta_1$ and $\theta_2$ of the
displacement operation
\begin{align}
D(\theta_1,\theta_2)= \exp (i \theta_2 \mathcal Q - i \theta_1
  \mathcal P  )\;,
\end{align}
acting on a probe state. It was shown in Refs.~
\cite{PhysRevLett.87.270404,PhysRevA.87.012107} that by using a two-mode entangled probe, one can estimate the displacement to arbitrary high accuracy. The probe is a symmetric two-mode squeezed thermal state. If the state is pure, it is known as a two-mode squeezed vacuum state, or an Einstein-Podolski-Rosen (EPR) state~\cite{weedbrook2012gaussian}. By symmetric we mean that the state has equal squeezing and noise in all quadratures. 

A measurement was proposed that can give an
arbitrarily precise estimate of both $\theta_1$ and $\theta_2$
simultaneously. This measurement, which resembles continuous variable
super-dense coding~\cite{PhysRevA.61.042302}, involves passing one
mode on an entangled probe to sense the displacement operation and
then jointly measuring it with an entangled ancilla. We call this
measurement the double-homodyne joint measurement~[see
Fig.~\ref{fig_measurement}(b)]. This extremely precise estimation scheme
was experimentally demonstrated in an optical
system~\cite{PhysRevLett.88.047904}.

Genoni \textit{et al.}~\cite{PhysRevA.87.012107} showed that for a symmetric two-mode squeezed state probe, in the limit of large entanglement, the
double-homodyne joint measurement approaches the ultimate precision
bounds calculated using the symmetric logarithmic derivative (SLD)
quantum Fisher information. However, for a general finite squeezing
level, there is a gap between the precision of the estimation from
dual homodyne measurement and the limit set by the right logarithmic
derivative (RLD) and SLD quantum Fisher information. This is not
surprising since in general we know that the RLD and SLD bounds are not tight \cite{szczykulska2016multi}. This raises two questions:
(i) Can we derive tight bounds for the precision? and (ii) Is there a
better measurement that will give a higher precision than the dual
homodyne measurement?

We address these questions for a general two-mode Gaussian probe. In
this work, we calculate the Holevo \Cramer-Rao (CR)
bound~\cite{holevobook,Holevo1976}, which is an asymptotically
achievable bound under some conditions \cite{yamagata2013,
  hayashi2008asymptotic,guctua2006local, kahn2009local}. However,
unlike the RLD and SLD bounds, computing the Holevo bound is in
general a hard problem because it involves an optimisation of a
nonlinear function over a space of Hermitian matrices. To date, it
has been solved in only a few simple cases. Providing the states
satisfy certain conditions, an explicit formula can be found for
Gaussian states \cite{holevobook,holevo1975some} or pure states
\cite{fujiwara1995quantum, matsumoto2002new}. Suzuki found a formula
in terms of the RLD and SLD CR bounds, for a qubit state parameterized
by two parameters~\cite{suzuki2016explicit}.

Previously, we performed this optimization for the special case when
the probe was a pure two-mode entangled state, and one mode
experiences an unknown displacement~\cite{Bradshaw2017}. When the
probe is mixed or if the channel is dissipative, then the space of the
optimisation problem is over infinite dimensional Hermitian
matrices. However, for Gaussian states, the probe and measurement can
be completely characterised by its first and second
moment~\cite{holevobook,Holevo1976}. This reduces the optimisation
space to four-dimensional positive semi-definite matrices which can be solved
efficiently using semi-definite programming
(SDP)~\cite{vandenberghe1996semidefinite}. Furthermore, the SDP and
its dual program provide a necessary and sufficient condition for
optimality of the solution, which can be verified analytically. 
Holevo solved the problem for mean estimation of Gaussian states 40 ago~\cite{holevobook,Holevo1976}. 
Our contribution is to recognise this as an SDP that can be solved
efficiently.

For the specific case of a symmetric two-mode squeezed state, we find
that the double-homodyne joint measurement is an optimal measurement
when the squeezing level is high enough such that the probe is
entangled. When the probe is separable, we find that the
double-homodyne joint measurement is sub-optimal. We propose a different
measurement scheme which is optimal.

In this paper, we provide a recipe for calculating the ultimate
precision of an unbiased estimate of displacement using a two-mode
Gaussian probe. We start with an introduction to multi-parameter local
quantum estimation in Sec.\ \ref{sec_background}. In Sec.\ \ref{sec_calculation}, we formulate the problem of displacement
estimation for two-mode Gaussian states in terms of an SDP. Section
\ref{sec_workedexample} gives an application of this formalism to the
symmetric two-mode squeezed state. Finally, we end with some
concluding remarks in Sec.\ \ref{sec_conclusion}.

\section{Multi-parameter local estimation}
\label{sec_background}
In classical parameter estimation theory, one starts with a random
variable $X$ that depends on some unknown parameter vector
$\theta=(\theta_1,\theta_2,\ldots,\theta_N)$ through a
conditional probability density function $f(x;\theta)$. The
random variable $X$ arises from the measurement of some state
$\rho(\theta$. From $X$, one can form a vector function
$\hat\theta=\hat\theta(X)$ that gives an unbiased estimate of
$\theta$. The goal is to find a precise estimate of theta. The bound
on how precise these unbiased estimator can be is determined by the
CR bound~\cite{cramer2016mathematical,rao1992information},
\begin{align}
  V_\theta[\hat{\theta}] \geq \frac{1}{I},
\end{align}
which relates the mean-square error (MSE) matrix
\begin{align}
  V_\theta[\hat{\theta}]_{jk} \coloneqq  \mathbb{E}[(\hat\theta_j-\theta_j)(\hat\theta_k-\theta_k)]
\end{align}
to the classical Fisher information matrix
\begin{align}
  I_{jk} \coloneqq -\mathbb{E}\left[\frac{\partial^2 }{\partial \theta_j \partial \theta_k}\log f\right]\;.
\end{align}
Under certain conditions, this bound can be asymptotically achieved by
the maximum likelihood estimator. We are interested in the sum of the
MSE, obtained by taking the trace of the MSE matrix $\Sigma \coloneqq\Tr{V_\theta[\hat{\theta}]}$.

Quantum parameter estimation theory \cite{paris2009quantum,Helstrom1969,PhysRevLett.72.3439,braunstein1996generalized} aims to determine the
ultimate precision with which certain parameters $\theta$ can be
determined from a quantum state $\rho_\theta$ that depends on
those parameters. This was developed by Helstrom~\cite{Helstrom1969,HELSTROM1967101,helstrom1974noncommuting}, Holevo~\cite{holevobook,Holevo1976} and others~\cite{Yuen1973,belavkin1976generalized} in the 1970s. There exists a whole family of
quantum Fisher information matrices, each of which gives
rise to its own CR bounds to the mean-square error
matrix~\cite{petz2011introduction}. However, none of these bounds are generally tight. Two commonly used
CR bounds are based on the SLD~\cite{HELSTROM1967101,Helstrom1969} and RLD~\cite{Yuen1973,belavkin1976generalized} Fisher information matrix.

The SLD operators $\SLD_j$ and RLD operators
$\RLD_j$ are obtained as solutions to the implicit
operator equations
\begin{align}
  \frac{\partial \rho}{\partial \theta_j}&=\frac{1}{2}\left(\rho \SLD_j+ \SLD_j \rho \right)\;\text{(SLD)}\\
  \frac{\partial \rho}{\partial \theta_j}&=\rho \RLD_j \;\text{(RLD)}\;.
\end{align}
The SLD operators are Hermitian but the RLD operators might not be
Hermitian. From the log-derivative operators, the SLD and RLD Fisher
information matrices are defined by
\begin{align}
  G^\text{(S)}_{jk} &\coloneqq \tr (\rho \frac{1}{2}(\SLD_j \SLD_k+\SLD_k
                      \SLD_j) )  \;\text{(SLD)},\\
  G^\text{(R)}_{jk} &\coloneqq \tr (\rho \RLD_j L^{\text{(R)}\dagger}_k )  \;\text{(RLD)},
\end{align}
from which we get the two CR bounds 
\begin{align}
  \Sigma &\geq \Tr{(G^{(S)})^{-1}} \eqqcolon  C^\text{(S)}, \\
  \Sigma &\geq \Tr{\Re( G^{(R)})^{-1}}+  \text{TrAbs}\left\{\Im( G^{(R)})^{-1}\right\} \eqqcolon  C^\text{(R)}, 
\end{align}
where $\text{TrAbs}\{X\}$ is the sum of the absolute values of the
eigenvalues of a matrix $X$. The SLD CR bound, $C^\text{(S)}$ gives the
optimal precision in estimating each parameter separately. However, for multi-parameter estimation, if optimal measurements for measuring each parameter separately do not commute (which is usually the case), then the SLD bound is not attainable. 
The RLD bound, $C^\text{(R)}$ is also in
general not attainable. However, when $\RLD$ is Hermitian,
$C^\text{(R)}$ provides an achievable bound for the joint
estimates \cite{fujiwara1994linear,fujiwara1999estimation,fujiwara1994multi}. In general, there is no hierarchy between $C^\text{(S)}$
and  $C^\text{(R)}$. 

Holevo unified these two bounds through the Holevo CR
bound~\cite{holevobook,Holevo1976}. This bound is achieved in the asymptotic limit of a joint measurement over infinite copies of the state~\cite{yamagata2013}. The Holevo CR bound is always greater or equal to $C^\text{(S)}$ and $C^\text{(R)}$. The bound
involves a minimization over $X=(X_1,X_2,\ldots,X_N)$ where $X_j$ are
Hermitian operators that satisfy the unbiased conditions
\begin{align}
\label{eq_xcon1}
\tr(\rho X_j)&=0, \\
\label{eq_xcon2}
\tr(\frac{\partial \rho_\theta}{\partial\theta_j} X_k)&=\delta_{jk}.
\end{align}
The Holevo CR bound is
\begin{align}
\label{eq_hol2}
\mathcal{V} \geq  \min_{X} \text{Tr}\left\{ Z_\theta[X]\right\}
  +\text{TrAbs} \left\{\Im Z_\theta[X]\right\}  \coloneqq C^\text{(H)}\;,
\end{align}
where
\begin{align}
\label{eq_zmat}
Z_\theta[X]_{jk} \coloneqq \tr \left( \rho X_j X_k \right)\;.
\end{align}
Holevo derived this bound in his original work~\cite{holevobook,Holevo1976}, but the bound in this form was introduced by Nagaoka~\cite{nagaoka2005new}.
A major obstacle preventing the more widespread use of the Holevo CR
bound is that unlike the RLD and SLD bounds, which can be calculated
directly, the Holevo bound involves a nontrivial optimisation
problem.

\section{Holevo bound for mean value estimation with Gaussian probes}
\label{sec_calculation}

When the probe is Gaussian, Holevo's bound can be simplified. It can
be formulated in terms of the first and second moments of the probe
state only. In this section, we summarise Holevo's result on mean value
estimation of Gaussian probes. For the proofs and technicalities of
these results, we recommend the interested reader to consult Holevo's
original work~\cite{holevobook,Holevo1976}. 

\subsection{Holevo's bound}
\label{sec_holevobound}

We want to estimate two parameters
$\theta_1$ and $\theta_2$ that are imprinted on the displacement of a
two-mode Gaussian state. Extension to more parameters or mode are
straight forward (see Appendix C). To arrive at Holevo's result we need to
introduce some notations.

For any $z=\begin{bmatrix}y_1&x_1&y_2&x_2\end{bmatrix}^\intercal$ in a four-dimensional real
vector space $Z$, let
\begin{align}
  \mathcal{R}(z) = x_1 \mathcal{P}_1 + y_1 \mathcal{Q}_1 + x_2 \mathcal{P}_2 + y_2 \mathcal{Q}_2\;,
\end{align}
where $\mathcal{P}_j$ and $\mathcal{Q}_j$ are the usual quadrature operators for the $j$-th mode in quantum
optics. $\mathcal{R}(z)$ are called canonical observables, and the
canonical commutation relation becomes
\begin{align}
\label{eq_commutation}
  [\mathcal{R}(z), \mathcal{R}(z')] = i \Delta(z,z')\;,
\end{align}
where
\begin{align}
 \Delta(z,z') = x'_1 y_1 - x_1 y'_1 + x'_2 y_2 - x_2 y'_2  
\end{align}
is a skew-symmetric bilinear form. By the Baker-Campbell-Hausdorff formula, we have an equivalent
representation of the canonical commutation relation as
\begin{align}
  \mathcal V(z) \mathcal V(z') = \exp\left( \frac{i}{2} \Delta(z,z')
  \right) \mathcal V(z+z')\;,
\end{align}
where $\mathcal V(z)=\e^{i \, \mathcal R(z)}$ is the Weyl
operator. The characteristic function of a state $\mathcal S$ is then
defined through $\mathcal V(z)$ as
$\chi_z[\mathcal S]=\tr{\mathcal S \mathcal V(z)}$. This is the
inverse-Weyl or Wigner transform that maps an operator in the Hilbert
space to some square-integrable function in $Z$. We say $\mathcal S$ is Gaussian
if the state is completely characterized by its first and second
moments~\cite{weedbrook2012gaussian}:
\begin{align}
  \chi_z[\mathcal S] =\exp\left[i\, m(z) - \frac{1}{2} \alpha(z,z) \right]\;,
\end{align}
where
\begin{align}
\label{eq_mz}
  m(z) &= \tr (\mathcal S \mathcal R(z) )\\
\label{eq_alpha}
  \alpha(z,z') &= \frac{1}{2} \tr(\mathcal S \{\mathcal
                 R(z)-m(z),\mathcal R(z')-m(z')\} )\;
\end{align}
and $\{\mathcal A,\mathcal B\}=\mathcal A \mathcal B+\mathcal B
\mathcal A$. The mean value function $m$ is a function of the
unknown parameters through
\begin{align}
  m(z) = \theta_1 m_1(z) + \theta_2 m_2(z)\;.
\end{align}
The correlation function $\alpha$ is an inner
product on $Z$, which defines a Euclidean space $(Z,\alpha)$. Now
let $\mathcal{D}$ be the associated operator of the form $\Delta$ in
$(Z,\alpha)$,
\begin{align}
  \Delta(z,z') = \alpha(z, \mathcal{D} z')\; \forall \;z,z' \in Z\;.
\end{align}
Define $m_j \in Z$ by $m_j(z) = \alpha(m_j,z)$. 

Holevo's CR bound is 
\begin{align}
\label{eqn:hol_bound}
 \Sigma \geq   \inf_{\mathscr{F}}{ \Tr{F^{-1}}} \eqqcolon \Sigma_* 
\end{align}
where $F$ is a $2 \times 2$ matrix with components
\begin{equation}
\label{eqn:Fmatrix}
F_{jk}=\alpha(m_j,\mathscr{F}m_k)
\end{equation}
and the infimum is taken over all real symmetric operators
$\mathscr{F}$ in $Z$, such that the complex extension of $\mathscr{F}$
satisfies
\begin{equation}
\label{eq_conditions}
0\le (1+\tfrac{1}{2}i\mathcal{D})\mathscr{F}(1+\tfrac{1}{2}i\mathcal{D})\le(1+\tfrac{1}{2}i\mathcal{D})
\end{equation} 
in the complexification of the Euclidean space $(Z,\alpha)$. $\mathcal
A\leq \mathcal B$
denotes $\alpha(z, \mathcal A z)\leq \alpha(z,\mathcal B z) $ for all $z\in Z$. Since
$1+\frac{1}{2} i \mathcal{D}$ is positive definite,
constraint~(\ref{eq_conditions}) is
equivalent to
\begin{equation}
\label{eq_conditions2}
0\le \mathscr{F}\leq  (1+\tfrac{1}{2}i\mathcal{D})^{-1}.
\end{equation} 

\subsection{Optimal measurement}
\label{sec_optimal_measure}
For estimating the mean of Gaussian probes, Holevo showed that the bound can be attained by a Gaussian measurement. Let $\mathscr{F}_*$ be the operator in
$Z$ that furnishes the minimum in (\ref{eqn:hol_bound}) and $F_*$ be
the corresponding matrix in (\ref{eqn:Fmatrix}). The optimal estimator
are given by the observables $\mathcal R(z_j^*)$ where
\begin{align}
\label{zstar}
  \begin{bmatrix}
    z_1^*\\z_2^*
  \end{bmatrix}=
   F^{-1}_*  \begin{bmatrix}
 \mathscr{F}_* m_1\\ \mathscr{F}_* m_2
  \end{bmatrix}\;.
\end{align}
$\mathcal R(z_1^*)$ and $\mathcal R(z_2^*)$ can be measured
simultaneously to attain precision $\Sigma_*$.

\subsection{Matrix representation}
\label{sec_matrix_rep}

The optimisation problem for computing Holevo's bound can be expressed
as a semi-definite program. This can be clearly seen if we introduce four
vectors $\{e_1,e_2,e_3,e_4\}$ that forms an orthonormal basis in the Euclidean
space $(Z,\alpha)$ such that $\alpha(e_j,e_k)=\delta_{jk}$ and
introduce
\begin{align}
  \mathbb{D}_{jk} &\coloneqq \alpha(e_j, \mathcal{D} e_k)\;
\end{align}
and
\begin{align}
  \mathbb{M}_{jk} &\coloneqq \alpha(m_j,e_k),\\
  \mathbb{F}_{jk} &\coloneqq \alpha(e_j, \mathscr{F} e_k),
\end{align}
so that
\begin{align}
F_{jk} &=\alpha(m_j,\mathscr{F}m_k) \\
  &=\sum_{mn} \alpha(m_j,e_m)\,\alpha(e_m , \mathscr{F} e_n)\,
    \alpha(e_n,m_k)\\
&=\left( \mathbb{M} \mathbb{F} \mathbb{M}^\intercal \right)_{jk}\;
\end{align}
Let $\mathbb{S}^n$ be the set of all $n\times n$ real symmetric matrices. Holevo's bound is obtained as a solution
to the following program:
\begin{program}{Holevo's bound}
\begin{align}
  \label{sdp_hol}
  \Sigma_* = &\min_{\mathbb{F} \in \mathbb{S}^4} \Tr{F^{-1}} \\
 \textnormal{subject to }\;  &0\le \mathbb{F}\leq \mathbb{C}\;,
\end{align}
\end{program}
where  $F= \mathbb{M} \mathbb{F} \mathbb{M}^\intercal$ and
 $\mathbb{C}\coloneqq (1+\tfrac{1}{2}i\mathbb{D})^{-1}$.
This is recognised as an SDP (see Appendix \ref{sec_app_sdp}) that can be solved
efficiently using standard numerical techniques.

\section{Worked example: Symmetric two-mode squeezed state}
\label{sec_workedexample}
We illustrate the computation of Holevo's bound through a specific
example. We start with a mixed two-mode squeezed state $\rho_0=S_2(r)
(\rho_\text{th}(v)\otimes \rho_\text{th}(v)) S_2^\dagger(r)$ as our probe where
\begin{align}
  \rho_\text{th}(v)= \frac{2}{(1+2v)}\sum_{n}\left(\frac{2v-1}{2v+1} \right)^{n}\ket{n}\bra{n}
\end{align}
is a thermal state with mean photon number
$v-\frac{1}{2}$ and quadrature variance $\alpha(z,z)=v$. The vacuum state corresponds to $v=\frac{1}{2}$. The ket $\ket{n}$ is the Fock state with $n$ photons, and
\begin{align} 
S_2(r) \coloneqq  \exp\left(r a_1 a_2 - r a_1^\dagger a_2^\dagger \right)  
\end{align}
is the two-mode squeezing operator where $a_j$ and $a_j^\dagger$ are
the $j$-th mode annihilation and creation operators with commutation
relation $[a,a^\dagger]=1$. Having prepared the probe $\rho_0$, we send one
mode through a displacement
\begin{align}
  D(\theta_1,\theta_2) \coloneqq \exp \left( i \theta_2 \mathcal Q_1 -
  i \theta_1 \mathcal P_1\right)
\end{align}
to get $\rho_\theta$, where $\theta_1$ and $\theta_2$ are the two unknown parameters that we
wish to determine. In what follows, we shall compute the
Holevo bound and present a measurement that achieves this bound. We
then compare this bound with the RLD and SLD bounds.

\subsection{Problem formulation}
Having the state $\rho_\theta$, we can already write its
characteristic function and find Holevo's bound directly. But, instead,
we choose to perform a unitary transformation to
decouple the two modes of the probe. The transformation we perform is
\begin{align}
  U=\exp \left(\frac{\pi}{4} (a_1^\dagger a_2- a_1 a_2^\dagger)\right),
\end{align}
which corresponds to interfering the two modes on a 50:50 beam splitter.
This extra step is not necessary
but is done for convenience so that the intermediate expressions in
computing the bound become less cumbersome. This of course will not
change the final result since the unitary operation can be considered part of the measurement. The
correlation function is
\begin{equation}
\alpha(z,z') = 
v \begin{bmatrix}
y_1\\x_1\\y_2\\x_2
 \end{bmatrix}^\intercal
 \begin{bmatrix}
\e^{-2r} & 0 & 0 & 0 \\
0 & \e^{2r} & 0 & 0 \\
0 & 0 & \e^{2r} & 0 \\
0 & 0 & 0 & \e^{-2r}
\end{bmatrix}
 \begin{bmatrix}
y'_1\\x'_1\\y'_2\\x'_2
 \end{bmatrix}
\end{equation}
and mean
\begin{align}
  m(z) =  \frac{1}{\sqrt 2}\begin{bmatrix}
\theta_1 \\ \theta_2\\ -\theta_1 \\ -\theta_2
 \end{bmatrix}^\intercal
 \begin{bmatrix}
y_1\\x_1\\y_2\\x_2
 \end{bmatrix}\;.
\end{align}
From this, the two vectors $m_1$ and $m_2$ in $Z$ are
\begin{align}
m_1&=\frac{1}{v \sqrt{2}}
\begin{bmatrix}
 \e^{2r}&0&-\e^{-2r}&0
\end{bmatrix}^\intercal, \\
m_2&=\frac{1}{v \sqrt{2}}
\begin{bmatrix}
 0&\e^{-2r}&0&-\e^{2r}
\end{bmatrix}^\intercal\;.
\end{align}
We now pick four orthonormal bases in $(\alpha,Z)$. Holevo's bound
does not depend on our choice of basis, any basis would do, and one
such basis is:
\begin{align}
  \begin{bmatrix}
    e_1&e_2&e_3&e_4
  \end{bmatrix}=
  \frac{1}{\sqrt{v}}
  \begin{bmatrix}
    \e^{r}&0&0&0\\
    0&\e^{-r}&0&0\\
    0&0&\e^{-r}&0\\
    0&0&0&\e^{r}
  \end{bmatrix}\;.
\end{align}
In this basis,
\begin{align}
  \mathbb{M} =\frac{1}{\sqrt {2v}}
  \begin{bmatrix}
    \e^r&0&-e^{-r}&0\\
    0&\e^{-r}&0&-e^{r}
  \end{bmatrix}
\end{align}
and
\begin{align}
  \mathbb{D}=\frac{1}{v} \begin{bmatrix}
0 & 1 & 0 & 0 \\
-1 & 0 & 0 & 0 \\
0 & 0 & 0 & 1 \\
0 & 0 & -1 & 0
\end{bmatrix}.
\end{align}

We show in Appendix \ref{sec_app1} that the solution to the SDP
program~(\ref{sdp_hol}) is 
\begin{equation}
\label{eq_bound}
\Sigma_*=
\begin{cases}
\frac{4v^2-1}{2v \cosh 2r-1}   & \text{if } r < r_0\\
4v \e^{-2r} & \text{if } r \ge r_0 
\end{cases}
\end{equation}
where $r_0=\frac{1}{2}\log(2v)$ and an optimal $\mathbb{F}_*$
attaining this is
\begin{align*}
\mathbb{F}_* &=
\frac{2v}{4v^2-1}  \begin{bmatrix}
    2v-\e^{-2r}&0&0&0\\
    0&2v-\e^{2r}&0&0\\
    0&0&2v-\e^{2r}&0\\
    0&0&0&2v-\e^{-2r}
  \end{bmatrix}
\end{align*}
for $r<r_0$ and
\begin{align}
\mathbb{F}_* &=
  \begin{bmatrix}
    1&0&0&0\\
    0&0&0&0\\
    0&0&0&0\\
    0&0&0&1
  \end{bmatrix}
\end{align}
for $r\ge r_0$.

\subsection{Optimal measurements that attains the bound}
\label{sec_measurements}

\begin{figure}
\includegraphics[width=8cm]{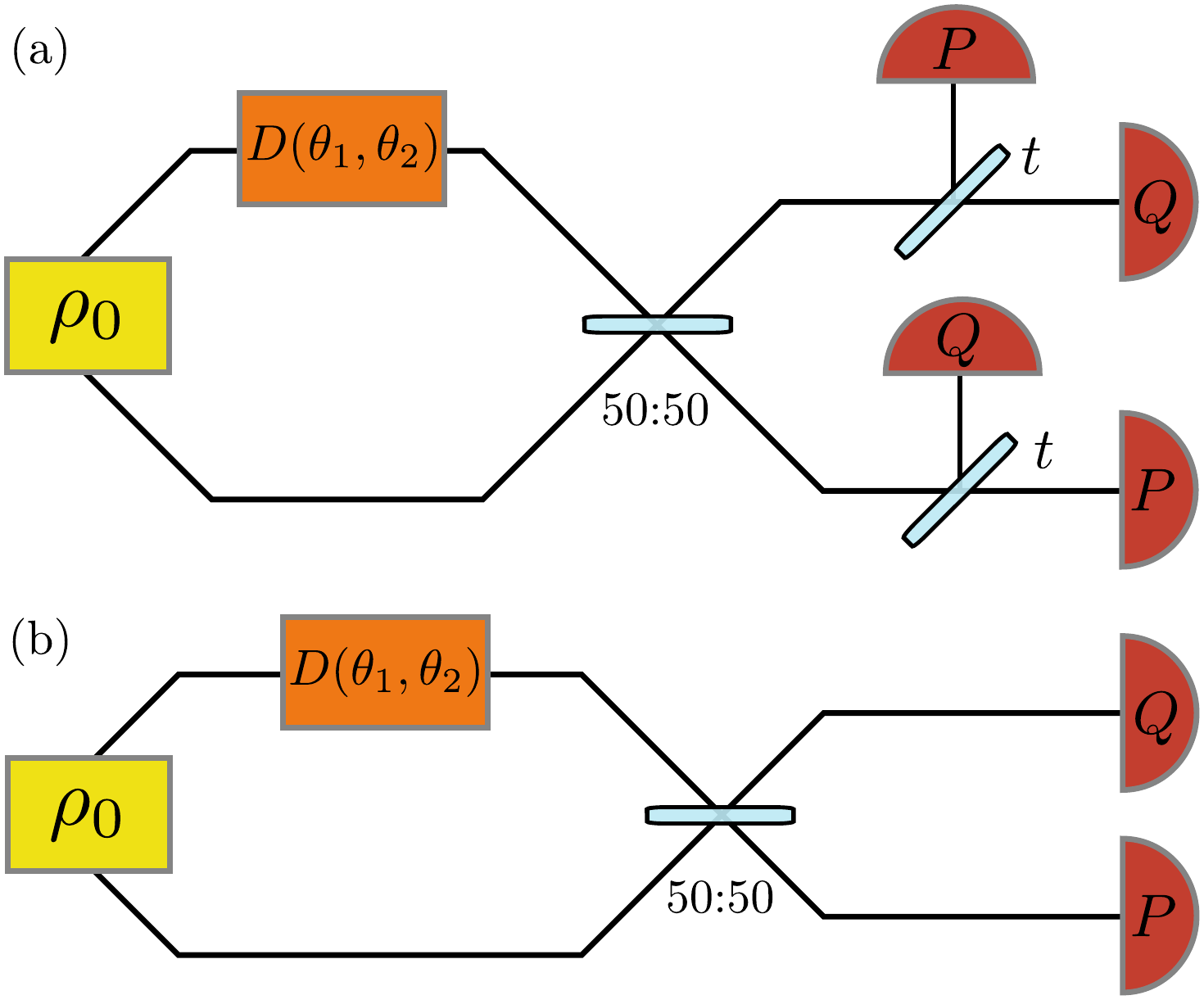}
\caption{The probe state $\rho_0$, a two-mode squeezed state, undergoes an unknown displacement $D(\theta_1,\theta_2)$. This figure shows the optimal measurement for estimating the displacement. (a) The optimal measurement to perform when $r< r_0$ is a double-unbalanced-heterodyne joint measurement. The two modes a mixed with a 50:50 beam splitter. Each output of the beam splitter then passes through another beam splitter with transmission $t$ given by~(\ref{eq_t}). Homodyne measurements of the $\mathcal P$ and $\mathcal Q$ quadratures are performed on the outputs of the beam splitters. (b) The optimal measurement to perform when $r\ge r_0$ is a double-homodyne joint measurement. The two modes are mixed with a 50:50 beam splitter. A homodyne measurement of the $\mathcal P$ quadrature is performed on one output of the beam splitter, and a homodyne measurement of the $\mathcal Q$ quadrature is performed on the other. }
\label{fig_measurement}
\end{figure}

To find the optimal measurement achieving $\Sigma_*$, we substitute
the solution for $\mathbb{F}_*$ into (\ref{zstar}) to obtain $z_j^*$. For $r<r_0$
\begin{align}
  z_1^* &= \sqrt{2} \begin{bmatrix}t&0&t-1&0\end{bmatrix}^\intercal,\\
  z_2^* &= \sqrt{2} \begin{bmatrix}0&1-t&0&-t\end{bmatrix}^\intercal,
\end{align}
where 
\begin{equation}
\label{eq_t}
t=\frac{2v e^{2r}-1}{4v\cosh 2r-2}. 
\end{equation} 
The observable
corresponding to this is
\begin{align}
\label{zstar_eg}
\begin{split}
  \mathcal R(z_1^*) &= \sqrt{2} t \mathcal Q_1 - \sqrt{2}(1-t)
                      \mathcal Q_2, \\
  \mathcal R(z_2^*) &= \sqrt{2} (1-t) \mathcal P_1 - \sqrt{2}t
                      \mathcal P_2\;,
\end{split}
\end{align}
whose physical realisation is shown in Fig.~\ref{fig_measurement}(a).

For $r\geq r_0$, we have
\begin{align}
  z_1^* &= \sqrt{2} \begin{bmatrix}1&0&0&0\end{bmatrix}^\intercal, \\
  z_2^* &= \sqrt{2} \begin{bmatrix}0&0&0&-1\end{bmatrix}^\intercal\;,
\end{align}
which is a special case of Eq.~(\ref{zstar_eg}) with $t=1$. The observables corresponding to these vectors are then
\begin{align}
  \mathcal R(z_1^*) &= \sqrt{2}  \mathcal Q_1, \\
  \mathcal R(z_2^*) &= -\sqrt{2} \mathcal P_2,
\end{align}
which is realized by the setup in Fig.~\ref{fig_measurement}(b). The two vectors $z_1^*$ and $z_2^*$ provide an unbiased estimator as can be checked by
noticing that $m(z_j^*)=\tr(\rho \mathcal R(z_j^*))= \theta_j$.

\subsection{Discussions}

\begin{figure}
\includegraphics[width=8.6cm]{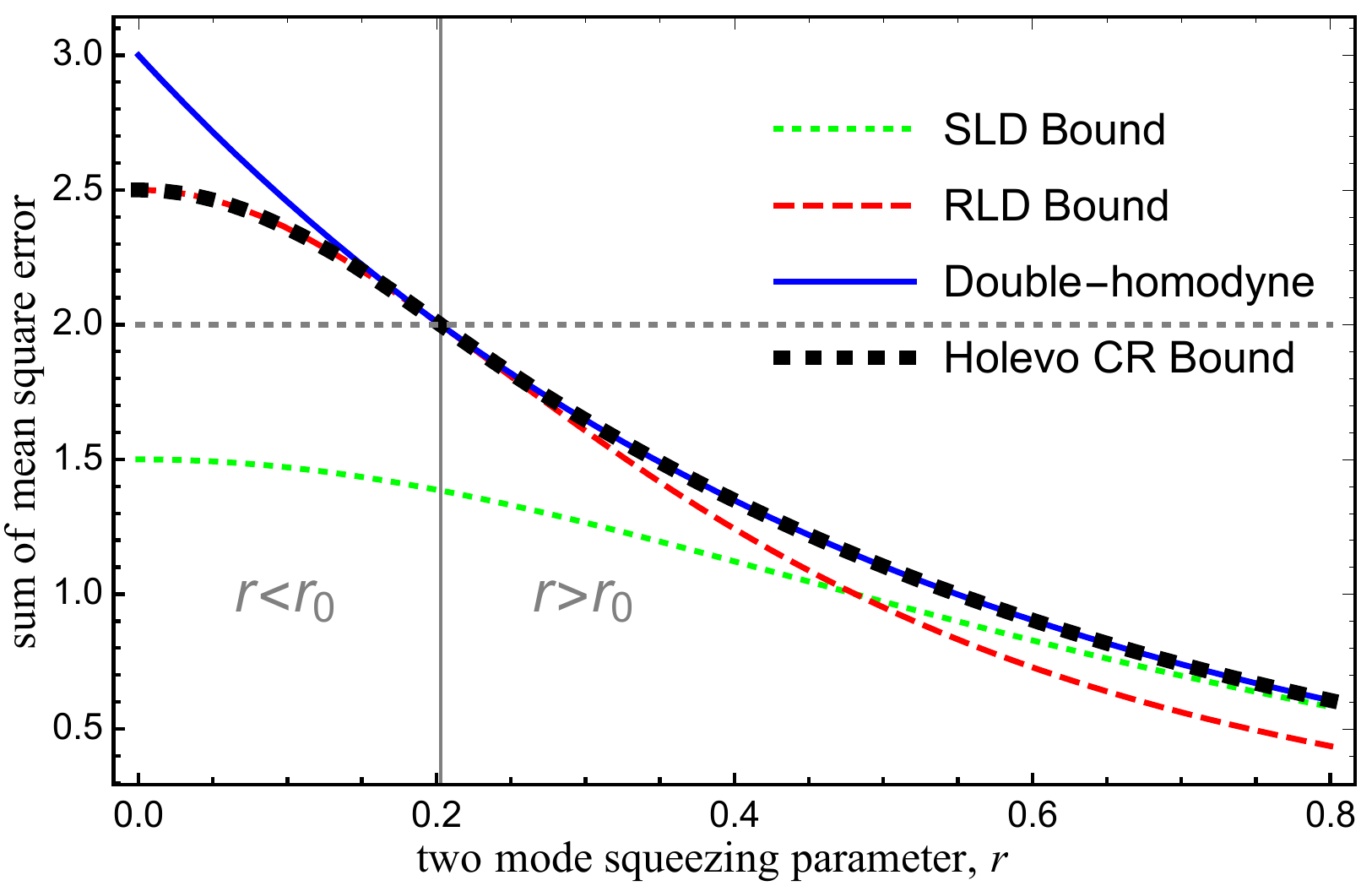}
\caption{Plot of symmetric logarithmic derivative (SLD), right logarithmic derivative (RLD), and Holevo \Cramer-Rao (CR) bounds, and double-homodyne joint measurement sum of mean-square error (MSE) for $v=0.75$. The vertical line corresponds to $r_0$. For $r\ge r_0$, the Holevo CR bound is equal to the sum of MSE of the double-homodyne joint measurement. For $r\le r_0$, the Holevo CR bound is equal to the RLD bound and the sum of MSE for the double-unbalanced-heterodyne joint measurement. The horizontal line corresponds to the best sum of MSE when using a single-mode Gaussian probe. }
\label{fig_plot1}
\end{figure}

Figure \ref{fig_plot1} shows the SLD and RLD CR bounds from Refs.~\cite{PhysRevA.87.012107, Gao2014}, our Holevo CR bound
Eq.~(\ref{eq_bound}), and the sum of MSE for a double-homodyne joint
measurement. The Holevo CR bound is greater than or equal to the RLD
and SLD CR bounds. When $r\ge r_0$, the sum of MSE for the
double-homodyne joint measurement is equal to the Holevo CR
bound. When $r \le r_0$, the Holevo CR bound is equal to the RLD CR
bound. The double-unbalanced-heterodyne joint measurement outperforms
the double-homodyne joint measurement in this case, giving a sum of
MSE equal to the RLD and Holevo CR bounds.  When $r>r_0$, the
double-unbalanced-heterodyne joint measurement is impossible,
requiring a beam splitter transmission greater than 1 [from
Eq.~(\ref{eq_t})].

Interestingly, we note that $r_0$ is the threshold beyond which the
probe becomes entangled as can be checked using Duan's inseparability
criterion~\cite{duan2000inseparability}. At $r=r_0$, the sum of MSE is
exactly $2$, which turns out to be the same as one get by doing a heterodyne
measurement on a single-mode coherent state probe. This is the best
one can do when restricted to single-mode Gaussian probes. Regardless
of whether the probe is entangled or not, the optimal measurement
scheme requires mixing the two modes on a 50:50 beam splitter, after
which we end up with two uncorrelated states. If the
probe state was originally entangled, the states after the 50:50
beam splitter will have a quadrature variance below the vacuum noise, while if
the original state is separable, all quadrature variances will always
be greater than the vacuum noise.

The double-unbalanced-heterodyne measurement can be seen as obtaining
two independent estimates for each displacement parameter and then
making an optimal estimate from these. As $t$ varies, the precision
of one estimate decreases at the expense of a better precision
for the second estimate. Suppose the system is entirely classical, and we have a classical state with covariances of $\mathcal{P}$ and $\mathcal{Q}$ the same as the quantum state. Because the system is classical, $\mathcal{P}$ and
$\mathcal{Q}$ can be measured simultaneously without an additional noise penalty imposed by quantum mechanics. In this case, the
double-unbalanced-heterodyne would outperform the dual-homodyne
measurement as we get two independent estimates for $\theta_1$ and two
independent estimates for $\theta_2$. However, for the quantum system, the
double-unbalanced-heterodyne measurement incurs a noise penalty due to
the vacuum noise coupling through the unused ports of the beam
splitters. There is a trade-off between a decreased precision due to
the vacuum noise, and an increased precision obtained from the
availability of an independent second estimate. When the measurement noise is greater
than the vacuum noise, the increase in precision we get from the
second estimate outweighs the loss of precision due to the vacuum
noise contaminating the first estimate. This is no longer true when
the measurement noise is smaller than the vacuum noise.

Even when the probe is separable, the optimal measurement still
requires a joint measurement of the two modes. Hence, perhaps
counter-intuitively, the optimal measurement is not separable despite
the probe being separable. Nevertheless, this is consistent with
previous work~\cite{gu2012observing}, where a joint measurement was
found to provide a higher mutual information than a separable
measurement. The performance advantage is attributed to the state
having a nonzero quantum discord, despite having no entanglement.

\section{Conclusion}
\label{sec_conclusion}
In conclusion, we provided a method to calculate the Holevo CR bound
for the estimation of the mean quadrature parameters of a two-mode
Gaussian state, by converting a problem to an SDP. An SDP can be
efficiently solved numerically. Additionally, conditions proving
optimality of an SDP solution exist, allowing for an analytical
solution to be verified. Our method can be easily extended to Gaussian states with any number of modes.

Using this method we were able to find an analytical solution for the
Holevo CR bound of the displacement on one mode of a symmetric two-mode squeezed
thermal state. A double-homodyne joint measurement is optimal if
the state is entangled, and a double-unbalanced-heterodyne joint measurement is
optimal if the state is separable.

\section*{Acknowledgements} 
This research is supported by the
Australian Research Council (ARC) under the Centre of Excellence for
Quantum Computation and Communication Technology (CE110001027). We would like to thank Nelly Ng for discussions and Jing Yan Haw for comments on the paper.

%
%
%
%

\bibliography{bib}

\begin{thebibliography}{47}%
\makeatletter
\providecommand \@ifxundefined [1]{%
 \@ifx{#1\undefined}
}%
\providecommand \@ifnum [1]{%
 \ifnum #1\expandafter \@firstoftwo
 \else \expandafter \@secondoftwo
 \fi
}%
\providecommand \@ifx [1]{%
 \ifx #1\expandafter \@firstoftwo
 \else \expandafter \@secondoftwo
 \fi
}%
\providecommand \natexlab [1]{#1}%
\providecommand \enquote  [1]{``#1''}%
\providecommand \bibnamefont  [1]{#1}%
\providecommand \bibfnamefont [1]{#1}%
\providecommand \citenamefont [1]{#1}%
\providecommand \href@noop [0]{\@secondoftwo}%
\providecommand \href [0]{\begingroup \@sanitize@url \@href}%
\providecommand \@href[1]{\@@startlink{#1}\@@href}%
\providecommand \@@href[1]{\endgroup#1\@@endlink}%
\providecommand \@sanitize@url [0]{\catcode `\\12\catcode `\$12\catcode
  `\&12\catcode `\#12\catcode `\^12\catcode `\_12\catcode `\%12\relax}%
\providecommand \@@startlink[1]{}%
\providecommand \@@endlink[0]{}%
\providecommand \url  [0]{\begingroup\@sanitize@url \@url }%
\providecommand \@url [1]{\endgroup\@href {#1}{\urlprefix }}%
\providecommand \urlprefix  [0]{URL }%
\providecommand \Eprint [0]{\href }%
\providecommand \doibase [0]{http://dx.doi.org/}%
\providecommand \selectlanguage [0]{\@gobble}%
\providecommand \bibinfo  [0]{\@secondoftwo}%
\providecommand \bibfield  [0]{\@secondoftwo}%
\providecommand \translation [1]{[#1]}%
\providecommand \BibitemOpen [0]{}%
\providecommand \bibitemStop [0]{}%
\providecommand \bibitemNoStop [0]{.\EOS\space}%
\providecommand \EOS [0]{\spacefactor3000\relax}%
\providecommand \BibitemShut  [1]{\csname bibitem#1\endcsname}%
\let\auto@bib@innerbib\@empty
\bibitem [{\citenamefont {Robertson}(1929)}]{PhysRev.34.163}%
  \BibitemOpen
  \bibfield  {author} {\bibinfo {author} {\bibfnamefont {H.~P.}\ \bibnamefont
  {Robertson}},\ }\bibfield  {title} {\enquote {\bibinfo {title} {The
  uncertainty principle},}\ }\href {http://dx.doi.org/10.1103/PhysRev.34.163}
  {\bibfield  {journal} {\bibinfo  {journal} {Phys. Rev.}\ }\textbf {\bibinfo
  {volume} {34}},\ \bibinfo {pages} {163--164} (\bibinfo {year}
  {1929})}\BibitemShut {NoStop}%
\bibitem [{\citenamefont {Weedbrook}\ \emph {et~al.}(2012)\citenamefont
  {Weedbrook}, \citenamefont {Pirandola}, \citenamefont {Garc\'{\i}a-Patr\'on},
  \citenamefont {Cerf}, \citenamefont {Ralph}, \citenamefont {Shapiro},\ and\
  \citenamefont {Lloyd}}]{weedbrook2012gaussian}%
  \BibitemOpen
  \bibfield  {author} {\bibinfo {author} {\bibfnamefont {C.}~\bibnamefont
  {Weedbrook}}, \bibinfo {author} {\bibfnamefont {S.}~\bibnamefont
  {Pirandola}}, \bibinfo {author} {\bibfnamefont {R.}~\bibnamefont
  {Garc\'{\i}a-Patr\'on}}, \bibinfo {author} {\bibfnamefont {N.~J.}\
  \bibnamefont {Cerf}}, \bibinfo {author} {\bibfnamefont {T.~C.}\ \bibnamefont
  {Ralph}}, \bibinfo {author} {\bibfnamefont {J.~H.}\ \bibnamefont {Shapiro}},
  \ and\ \bibinfo {author} {\bibfnamefont {S.}~\bibnamefont {Lloyd}},\
  }\bibfield  {title} {\enquote {\bibinfo {title} {Gaussian quantum
  information},}\ }\href {http://dx.doi.org/10.1103/RevModPhys.84.621}
  {\bibfield  {journal} {\bibinfo  {journal} {Rev. Mod. Phys.}\ }\textbf
  {\bibinfo {volume} {84}},\ \bibinfo {pages} {621--669} (\bibinfo {year}
  {2012})}\BibitemShut {NoStop}%
\bibitem [{\citenamefont {Adesso}\ \emph {et~al.}(2014)\citenamefont {Adesso},
  \citenamefont {Ragy},\ and\ \citenamefont {Lee}}]{adesso2014continuous}%
  \BibitemOpen
  \bibfield  {author} {\bibinfo {author} {\bibfnamefont {G.}~\bibnamefont
  {Adesso}}, \bibinfo {author} {\bibfnamefont {S.}~\bibnamefont {Ragy}}, \ and\
  \bibinfo {author} {\bibfnamefont {A.~R.}\ \bibnamefont {Lee}},\ }\bibfield
  {title} {\enquote {\bibinfo {title} {Continuous variable quantum information:
  Gaussian states and beyond},}\ }\href
  {http://dx.doi.org/10.1142/S1230161214400010} {\bibfield  {journal} {\bibinfo
   {journal} {Open Syst. Inf. Dyn.}\ }\textbf {\bibinfo {volume} {21}},\
  \bibinfo {pages} {1440001} (\bibinfo {year} {2014})}\BibitemShut {NoStop}%
\bibitem [{\citenamefont {Giovannetti}\ \emph {et~al.}(2004)\citenamefont
  {Giovannetti}, \citenamefont {Lloyd},\ and\ \citenamefont
  {Maccone}}]{Giovannetti1330}%
  \BibitemOpen
  \bibfield  {author} {\bibinfo {author} {\bibfnamefont {V.}~\bibnamefont
  {Giovannetti}}, \bibinfo {author} {\bibfnamefont {S.}~\bibnamefont {Lloyd}},
  \ and\ \bibinfo {author} {\bibfnamefont {L.}~\bibnamefont {Maccone}},\
  }\bibfield  {title} {\enquote {\bibinfo {title} {Quantum-enhanced
  measurements: Beating the standard quantum limit},}\ }\href
  {http://dx.doi.org/10.1126/science.1104149} {\bibfield  {journal} {\bibinfo
  {journal} {Science}\ }\textbf {\bibinfo {volume} {306}},\ \bibinfo {pages}
  {1330--1336} (\bibinfo {year} {2004})}\BibitemShut {NoStop}%
\bibitem [{\citenamefont {D'Ariano}\ \emph {et~al.}(2001)\citenamefont
  {D'Ariano}, \citenamefont {Lo~Presti},\ and\ \citenamefont
  {Paris}}]{PhysRevLett.87.270404}%
  \BibitemOpen
  \bibfield  {author} {\bibinfo {author} {\bibfnamefont {G.~M.}\ \bibnamefont
  {D'Ariano}}, \bibinfo {author} {\bibfnamefont {P.}~\bibnamefont {Lo~Presti}},
  \ and\ \bibinfo {author} {\bibfnamefont {M.~G.~A.}\ \bibnamefont {Paris}},\
  }\bibfield  {title} {\enquote {\bibinfo {title} {Using entanglement improves
  the precision of quantum measurements},}\ }\href
  {http://dx.doi.org/10.1103/PhysRevLett.87.270404} {\bibfield  {journal}
  {\bibinfo  {journal} {Phys. Rev. Lett.}\ }\textbf {\bibinfo {volume} {87}},\
  \bibinfo {pages} {270404} (\bibinfo {year} {2001})}\BibitemShut {NoStop}%
\bibitem [{\citenamefont {Fujiwara}(2001)}]{fujiwara2001quantum}%
  \BibitemOpen
  \bibfield  {author} {\bibinfo {author} {\bibfnamefont {A.}~\bibnamefont
  {Fujiwara}},\ }\bibfield  {title} {\enquote {\bibinfo {title} {Quantum
  channel identification problem},}\ }\href
  {https://doi.org/10.1103/PhysRevA.63.042304} {\bibfield  {journal} {\bibinfo
  {journal} {Phys. Rev. A}\ }\textbf {\bibinfo {volume} {63}},\ \bibinfo
  {pages} {042304} (\bibinfo {year} {2001})}\BibitemShut {NoStop}%
\bibitem [{\citenamefont {Fischer}\ \emph {et~al.}(2001)\citenamefont
  {Fischer}, \citenamefont {Mack}, \citenamefont {Cirone},\ and\ \citenamefont
  {Freyberger}}]{fischer2001enhanced}%
  \BibitemOpen
  \bibfield  {author} {\bibinfo {author} {\bibfnamefont {D.~G.}\ \bibnamefont
  {Fischer}}, \bibinfo {author} {\bibfnamefont {H.}~\bibnamefont {Mack}},
  \bibinfo {author} {\bibfnamefont {M.~A.}\ \bibnamefont {Cirone}}, \ and\
  \bibinfo {author} {\bibfnamefont {M.}~\bibnamefont {Freyberger}},\ }\bibfield
   {title} {\enquote {\bibinfo {title} {Enhanced estimation of a noisy quantum
  channel using entanglement},}\ }\href
  {https://doi.org/10.1103/PhysRevA.64.022309} {\bibfield  {journal} {\bibinfo
  {journal} {Phys. Rev. A}\ }\textbf {\bibinfo {volume} {64}},\ \bibinfo
  {pages} {022309} (\bibinfo {year} {2001})}\BibitemShut {NoStop}%
\bibitem [{\citenamefont {Sasaki}\ \emph {et~al.}(2002)\citenamefont {Sasaki},
  \citenamefont {Ban},\ and\ \citenamefont {Barnett}}]{sasaki2002optimal}%
  \BibitemOpen
  \bibfield  {author} {\bibinfo {author} {\bibfnamefont {M.}~\bibnamefont
  {Sasaki}}, \bibinfo {author} {\bibfnamefont {M.}~\bibnamefont {Ban}}, \ and\
  \bibinfo {author} {\bibfnamefont {S.~M.}\ \bibnamefont {Barnett}},\
  }\bibfield  {title} {\enquote {\bibinfo {title} {Optimal parameter estimation
  of a depolarizing channel},}\ }\href
  {https://doi.org/10.1103/PhysRevA.66.022308} {\bibfield  {journal} {\bibinfo
  {journal} {Phys. Rev. A}\ }\textbf {\bibinfo {volume} {66}},\ \bibinfo
  {pages} {022308} (\bibinfo {year} {2002})}\BibitemShut {NoStop}%
\bibitem [{\citenamefont {Fujiwara}\ and\ \citenamefont
  {Imai}(2003)}]{fujiwara2003quantum}%
  \BibitemOpen
  \bibfield  {author} {\bibinfo {author} {\bibfnamefont {A.}~\bibnamefont
  {Fujiwara}}\ and\ \bibinfo {author} {\bibfnamefont {H.}~\bibnamefont
  {Imai}},\ }\bibfield  {title} {\enquote {\bibinfo {title} {Quantum parameter
  estimation of a generalized pauli channel},}\ }\href
  {https://doi.org/10.1088/0305-4470/36/29/314} {\bibfield  {journal} {\bibinfo
   {journal} {J. Phys. A: Math. Gen.}\ }\textbf {\bibinfo {volume} {36}},\
  \bibinfo {pages} {8093} (\bibinfo {year} {2003})}\BibitemShut {NoStop}%
\bibitem [{\citenamefont {Ballester}(2004)}]{ballester2004estimation}%
  \BibitemOpen
  \bibfield  {author} {\bibinfo {author} {\bibfnamefont {M.~A.}\ \bibnamefont
  {Ballester}},\ }\bibfield  {title} {\enquote {\bibinfo {title} {Estimation of
  unitary quantum operations},}\ }\href
  {https://doi.org/10.1103/PhysRevA.69.022303} {\bibfield  {journal} {\bibinfo
  {journal} {Phys. Rev. A}\ }\textbf {\bibinfo {volume} {69}},\ \bibinfo
  {pages} {022303} (\bibinfo {year} {2004})}\BibitemShut {NoStop}%
\bibitem [{\citenamefont {Rigovacca}\ \emph {et~al.}(2017)\citenamefont
  {Rigovacca}, \citenamefont {Farace}, \citenamefont {Souza}, \citenamefont
  {De~Pasquale}, \citenamefont {Giovannetti},\ and\ \citenamefont
  {Adesso}}]{rigovacca2017versatile}%
  \BibitemOpen
  \bibfield  {author} {\bibinfo {author} {\bibfnamefont {L.}~\bibnamefont
  {Rigovacca}}, \bibinfo {author} {\bibfnamefont {A.}~\bibnamefont {Farace}},
  \bibinfo {author} {\bibfnamefont {L.~A.~M.}\ \bibnamefont {Souza}}, \bibinfo
  {author} {\bibfnamefont {A.}~\bibnamefont {De~Pasquale}}, \bibinfo {author}
  {\bibfnamefont {V.}~\bibnamefont {Giovannetti}}, \ and\ \bibinfo {author}
  {\bibfnamefont {G.}~\bibnamefont {Adesso}},\ }\bibfield  {title} {\enquote
  {\bibinfo {title} {Versatile gaussian probes for squeezing estimation},}\
  }\href {https://doi.org/10.1103/PhysRevA.95.052331} {\bibfield  {journal}
  {\bibinfo  {journal} {Phys. Rev. A}\ }\textbf {\bibinfo {volume} {95}},\
  \bibinfo {pages} {052331} (\bibinfo {year} {2017})}\BibitemShut {NoStop}%
\bibitem [{\citenamefont {Ruppert}\ and\ \citenamefont
  {Filip}(2017)}]{ruppert2017light}%
  \BibitemOpen
  \bibfield  {author} {\bibinfo {author} {\bibfnamefont {L.}~\bibnamefont
  {Ruppert}}\ and\ \bibinfo {author} {\bibfnamefont {R.}~\bibnamefont
  {Filip}},\ }\bibfield  {title} {\enquote {\bibinfo {title} {Light-matter
  quantum interferometry with homodyne detection},}\ }\href
  {http://dx.doi.org/10.1364/OE.25.015456} {\bibfield  {journal} {\bibinfo
  {journal} {Opt. Express}\ }\textbf {\bibinfo {volume} {25}},\ \bibinfo
  {pages} {15456--15467} (\bibinfo {year} {2017})}\BibitemShut {NoStop}%
\bibitem [{\citenamefont {Duivenvoorden}\ \emph {et~al.}(2017)\citenamefont
  {Duivenvoorden}, \citenamefont {Terhal},\ and\ \citenamefont
  {Weigand}}]{duivenvoorden2017single}%
  \BibitemOpen
  \bibfield  {author} {\bibinfo {author} {\bibfnamefont {K.}~\bibnamefont
  {Duivenvoorden}}, \bibinfo {author} {\bibfnamefont {B.~M.}\ \bibnamefont
  {Terhal}}, \ and\ \bibinfo {author} {\bibfnamefont {D.}~\bibnamefont
  {Weigand}},\ }\bibfield  {title} {\enquote {\bibinfo {title} {Single-mode
  displacement sensor},}\ }\href {https://doi.org/10.1103/PhysRevA.95.012305}
  {\bibfield  {journal} {\bibinfo  {journal} {Phys. Rev. A}\ }\textbf {\bibinfo
  {volume} {95}},\ \bibinfo {pages} {012305} (\bibinfo {year}
  {2017})}\BibitemShut {NoStop}%
\bibitem [{\citenamefont {Genoni}\ \emph {et~al.}(2013)\citenamefont {Genoni},
  \citenamefont {Paris}, \citenamefont {Adesso}, \citenamefont {Nha},
  \citenamefont {Knight},\ and\ \citenamefont {Kim}}]{PhysRevA.87.012107}%
  \BibitemOpen
  \bibfield  {author} {\bibinfo {author} {\bibfnamefont {M.~G.}\ \bibnamefont
  {Genoni}}, \bibinfo {author} {\bibfnamefont {M.~G.~A.}\ \bibnamefont
  {Paris}}, \bibinfo {author} {\bibfnamefont {G.}~\bibnamefont {Adesso}},
  \bibinfo {author} {\bibfnamefont {H.}~\bibnamefont {Nha}}, \bibinfo {author}
  {\bibfnamefont {P.~L.}\ \bibnamefont {Knight}}, \ and\ \bibinfo {author}
  {\bibfnamefont {M.~S.}\ \bibnamefont {Kim}},\ }\bibfield  {title} {\enquote
  {\bibinfo {title} {Optimal estimation of joint parameters in phase space},}\
  }\href {http://dx.doi.org/10.1103/PhysRevA.87.012107} {\bibfield  {journal}
  {\bibinfo  {journal} {Phys. Rev. A}\ }\textbf {\bibinfo {volume} {87}},\
  \bibinfo {pages} {012107} (\bibinfo {year} {2013})}\BibitemShut {NoStop}%
\bibitem [{\citenamefont {Braunstein}\ and\ \citenamefont
  {Kimble}(2000)}]{PhysRevA.61.042302}%
  \BibitemOpen
  \bibfield  {author} {\bibinfo {author} {\bibfnamefont {S.~L.}\ \bibnamefont
  {Braunstein}}\ and\ \bibinfo {author} {\bibfnamefont {H.~J.}\ \bibnamefont
  {Kimble}},\ }\bibfield  {title} {\enquote {\bibinfo {title} {Dense coding for
  continuous variables},}\ }\href
  {http://dx.doi.org/10.1103/PhysRevA.61.042302} {\bibfield  {journal}
  {\bibinfo  {journal} {Phys. Rev. A}\ }\textbf {\bibinfo {volume} {61}},\
  \bibinfo {pages} {042302} (\bibinfo {year} {2000})}\BibitemShut {NoStop}%
\bibitem [{\citenamefont {Li}\ \emph {et~al.}(2002)\citenamefont {Li},
  \citenamefont {Pan}, \citenamefont {Jing}, \citenamefont {Zhang},
  \citenamefont {Xie},\ and\ \citenamefont {Peng}}]{PhysRevLett.88.047904}%
  \BibitemOpen
  \bibfield  {author} {\bibinfo {author} {\bibfnamefont {X.}~\bibnamefont
  {Li}}, \bibinfo {author} {\bibfnamefont {Q.}~\bibnamefont {Pan}}, \bibinfo
  {author} {\bibfnamefont {J.}~\bibnamefont {Jing}}, \bibinfo {author}
  {\bibfnamefont {J.}~\bibnamefont {Zhang}}, \bibinfo {author} {\bibfnamefont
  {C.}~\bibnamefont {Xie}}, \ and\ \bibinfo {author} {\bibfnamefont
  {K.}~\bibnamefont {Peng}},\ }\bibfield  {title} {\enquote {\bibinfo {title}
  {Quantum dense coding exploiting a bright einstein-podolsky-rosen beam},}\
  }\href {http://dx.doi.org/10.1103/PhysRevLett.88.047904} {\bibfield
  {journal} {\bibinfo  {journal} {Phys. Rev. Lett.}\ }\textbf {\bibinfo
  {volume} {88}},\ \bibinfo {pages} {047904} (\bibinfo {year}
  {2002})}\BibitemShut {NoStop}%
\bibitem [{\citenamefont {Szczykulska}\ \emph {et~al.}(2016)\citenamefont
  {Szczykulska}, \citenamefont {Baumgratz},\ and\ \citenamefont
  {Datta}}]{szczykulska2016multi}%
  \BibitemOpen
  \bibfield  {author} {\bibinfo {author} {\bibfnamefont {M.}~\bibnamefont
  {Szczykulska}}, \bibinfo {author} {\bibfnamefont {T.}~\bibnamefont
  {Baumgratz}}, \ and\ \bibinfo {author} {\bibfnamefont {A.}~\bibnamefont
  {Datta}},\ }\bibfield  {title} {\enquote {\bibinfo {title} {Multi-parameter
  quantum metrology},}\ }\href
  {http://dx.doi.org/10.1080/23746149.2016.1230476} {\bibfield  {journal}
  {\bibinfo  {journal} {Adv. Phys: X}\ }\textbf {\bibinfo {volume} {1}},\
  \bibinfo {pages} {621--639} (\bibinfo {year} {2016})}\BibitemShut {NoStop}%
\bibitem [{\citenamefont {Holevo}(2011)}]{holevobook}%
  \BibitemOpen
  \bibfield  {author} {\bibinfo {author} {\bibfnamefont {A.~S.}\ \bibnamefont
  {Holevo}},\ }\href@noop {} {\emph {\bibinfo {title} {Probabilistic and
  statistical aspects of quantum theory}}},\ Vol.~\bibinfo {volume} {1}\
  (\bibinfo  {publisher} {Springer Science \& Business Media},\ \bibinfo {year}
  {2011})\BibitemShut {NoStop}%
\bibitem [{\citenamefont {Holevo}(1976)}]{Holevo1976}%
  \BibitemOpen
  \bibfield  {author} {\bibinfo {author} {\bibfnamefont {A.~S.}\ \bibnamefont
  {Holevo}},\ }\enquote {\bibinfo {title} {Noncommutative analogues of the
  cram{\'e}r-rao inequality in the quantum measurement theory},}\ in\ \href
  {http://dx.doi.org/10.1007/BFb0077491} {\emph {\bibinfo {booktitle}
  {Proceedings of the Third Japan --- USSR Symposium on Probability Theory}}},\
  \bibinfo {editor} {edited by\ \bibinfo {editor} {\bibfnamefont
  {G.}~\bibnamefont {Maruyama}}\ and\ \bibinfo {editor} {\bibfnamefont {J.~V.}\
  \bibnamefont {Prokhorov}}}\ (\bibinfo  {publisher} {Springer Berlin
  Heidelberg},\ \bibinfo {address} {Berlin, Heidelberg},\ \bibinfo {year}
  {1976})\ pp.\ \bibinfo {pages} {194--222}\BibitemShut {NoStop}%
\bibitem [{\citenamefont {Yamagata}\ \emph {et~al.}(2013)\citenamefont
  {Yamagata}, \citenamefont {Fujiwara},\ and\ \citenamefont
  {Gill}}]{yamagata2013}%
  \BibitemOpen
  \bibfield  {author} {\bibinfo {author} {\bibfnamefont {K.}~\bibnamefont
  {Yamagata}}, \bibinfo {author} {\bibfnamefont {A.}~\bibnamefont {Fujiwara}},
  \ and\ \bibinfo {author} {\bibfnamefont {R.~D.}\ \bibnamefont {Gill}},\
  }\bibfield  {title} {\enquote {\bibinfo {title} {Quantum local asymptotic
  normality based on a new quantum likelihood ratio},}\ }\href
  {http://www.jstor.org/stable/23566545} {\bibfield  {journal} {\bibinfo
  {journal} {Ann. Stat.}\ }\textbf {\bibinfo {volume} {41}},\ \bibinfo {pages}
  {2197--2217} (\bibinfo {year} {2013})}\BibitemShut {NoStop}%
\bibitem [{\citenamefont {Hayashi}\ and\ \citenamefont
  {Matsumoto}(2008)}]{hayashi2008asymptotic}%
  \BibitemOpen
  \bibfield  {author} {\bibinfo {author} {\bibfnamefont {M.}~\bibnamefont
  {Hayashi}}\ and\ \bibinfo {author} {\bibfnamefont {K.}~\bibnamefont
  {Matsumoto}},\ }\bibfield  {title} {\enquote {\bibinfo {title} {Asymptotic
  performance of optimal state estimation in qubit system},}\ }\href
  {http://dx.doi.org/10.1063/1.2988130} {\bibfield  {journal} {\bibinfo
  {journal} {J. Math. Phys.}\ }\textbf {\bibinfo {volume} {49}},\ \bibinfo
  {pages} {102101} (\bibinfo {year} {2008})}\BibitemShut {NoStop}%
\bibitem [{\citenamefont {Gu{\c{t}}{\u{a}}}\ and\ \citenamefont
  {Kahn}(2006)}]{guctua2006local}%
  \BibitemOpen
  \bibfield  {author} {\bibinfo {author} {\bibfnamefont {M.}~\bibnamefont
  {Gu{\c{t}}{\u{a}}}}\ and\ \bibinfo {author} {\bibfnamefont {J.}~\bibnamefont
  {Kahn}},\ }\bibfield  {title} {\enquote {\bibinfo {title} {Local asymptotic
  normality for qubit states},}\ }\href
  {https://doi.org/10.1103/PhysRevA.73.052108} {\bibfield  {journal} {\bibinfo
  {journal} {Phys. Rev. A}\ }\textbf {\bibinfo {volume} {73}},\ \bibinfo
  {pages} {052108} (\bibinfo {year} {2006})}\BibitemShut {NoStop}%
\bibitem [{\citenamefont {Kahn}\ and\ \citenamefont
  {Gu{\c{t}}{\u{a}}}(2009)}]{kahn2009local}%
  \BibitemOpen
  \bibfield  {author} {\bibinfo {author} {\bibfnamefont {J.}~\bibnamefont
  {Kahn}}\ and\ \bibinfo {author} {\bibfnamefont {M.}~\bibnamefont
  {Gu{\c{t}}{\u{a}}}},\ }\bibfield  {title} {\enquote {\bibinfo {title} {Local
  asymptotic normality for finite dimensional quantum systems},}\ }\href
  {https://doi.org/10.1007/s00220-009-0787-3} {\bibfield  {journal} {\bibinfo
  {journal} {Commun. Math. Phys.}\ }\textbf {\bibinfo {volume} {289}},\
  \bibinfo {pages} {597--652} (\bibinfo {year} {2009})}\BibitemShut {NoStop}%
\bibitem [{\citenamefont {Holevo}(1975)}]{holevo1975some}%
  \BibitemOpen
  \bibfield  {author} {\bibinfo {author} {\bibfnamefont {A.~S.}\ \bibnamefont
  {Holevo}},\ }\bibfield  {title} {\enquote {\bibinfo {title} {Some statistical
  problems for quantum gaussian states},}\ }\href
  {https://doi.org/10.1109/TIT.1975.1055441} {\bibfield  {journal} {\bibinfo
  {journal} {IEEE T. Inform. Theory}\ }\textbf {\bibinfo {volume} {21}},\
  \bibinfo {pages} {533--543} (\bibinfo {year} {1975})}\BibitemShut {NoStop}%
\bibitem [{\citenamefont {Fujiwara}\ and\ \citenamefont
  {Nagaoka}(1995)}]{fujiwara1995quantum}%
  \BibitemOpen
  \bibfield  {author} {\bibinfo {author} {\bibfnamefont {A.}~\bibnamefont
  {Fujiwara}}\ and\ \bibinfo {author} {\bibfnamefont {H.}~\bibnamefont
  {Nagaoka}},\ }\bibfield  {title} {\enquote {\bibinfo {title} {Quantum fisher
  metric and estimation for pure state models},}\ }\href
  {https://doi.org/10.1016/0375-9601(95)00269-9} {\bibfield  {journal}
  {\bibinfo  {journal} {Phys. Lett. A}\ }\textbf {\bibinfo {volume} {201}},\
  \bibinfo {pages} {119--124} (\bibinfo {year} {1995})}\BibitemShut {NoStop}%
\bibitem [{\citenamefont {Matsumoto}(2002)}]{matsumoto2002new}%
  \BibitemOpen
  \bibfield  {author} {\bibinfo {author} {\bibfnamefont {K.}~\bibnamefont
  {Matsumoto}},\ }\bibfield  {title} {\enquote {\bibinfo {title} {A new
  approach to the cram{\'e}r-rao-type bound of the pure-state model},}\ }\href
  {https://doi.org/10.1088/0305-4470/35/13/307} {\bibfield  {journal} {\bibinfo
   {journal} {J. Phys. A: Math. Gen.}\ }\textbf {\bibinfo {volume} {35}},\
  \bibinfo {pages} {3111} (\bibinfo {year} {2002})}\BibitemShut {NoStop}%
\bibitem [{\citenamefont {Suzuki}(2016)}]{suzuki2016explicit}%
  \BibitemOpen
  \bibfield  {author} {\bibinfo {author} {\bibfnamefont {J.}~\bibnamefont
  {Suzuki}},\ }\bibfield  {title} {\enquote {\bibinfo {title} {Explicit formula
  for the holevo bound for two-parameter qubit-state estimation problem},}\
  }\href {http://dx.doi.org/10.1063/1.4945086} {\bibfield  {journal} {\bibinfo
  {journal} {J. Math. Phys.}\ }\textbf {\bibinfo {volume} {57}},\ \bibinfo
  {pages} {042201} (\bibinfo {year} {2016})}\BibitemShut {NoStop}%
\bibitem [{\citenamefont {Bradshaw}\ \emph {et~al.}(2017)\citenamefont
  {Bradshaw}, \citenamefont {Assad},\ and\ \citenamefont {Lam}}]{Bradshaw2017}%
  \BibitemOpen
  \bibfield  {author} {\bibinfo {author} {\bibfnamefont {M.}~\bibnamefont
  {Bradshaw}}, \bibinfo {author} {\bibfnamefont {S.~M.}\ \bibnamefont {Assad}},
  \ and\ \bibinfo {author} {\bibfnamefont {P.~K.}\ \bibnamefont {Lam}},\
  }\bibfield  {title} {\enquote {\bibinfo {title} {A tight cram{\'e}r--rao
  bound for joint parameter estimation with a pure two-mode squeezed probe},}\
  }\href {https://doi.org/10.1016/j.physleta.2017.06.024} {\bibfield  {journal}
  {\bibinfo  {journal} {Phys. Lett. A}\ }\textbf {\bibinfo {volume} {381}},\
  \bibinfo {pages} {2598 -- 2607} (\bibinfo {year} {2017})}\BibitemShut
  {NoStop}%
\bibitem [{\citenamefont {Vandenberghe}\ and\ \citenamefont
  {Boyd}(1996)}]{vandenberghe1996semidefinite}%
  \BibitemOpen
  \bibfield  {author} {\bibinfo {author} {\bibfnamefont {L.}~\bibnamefont
  {Vandenberghe}}\ and\ \bibinfo {author} {\bibfnamefont {S.}~\bibnamefont
  {Boyd}},\ }\bibfield  {title} {\enquote {\bibinfo {title} {Semidefinite
  programming},}\ }\href {https://doi.org/10.1137/1038003} {\bibfield
  {journal} {\bibinfo  {journal} {SIAM Rev.}\ }\textbf {\bibinfo {volume}
  {38}},\ \bibinfo {pages} {49--95} (\bibinfo {year} {1996})}\BibitemShut
  {NoStop}%
\bibitem [{\citenamefont {Cram{\'e}r}(2016)}]{cramer2016mathematical}%
  \BibitemOpen
  \bibfield  {author} {\bibinfo {author} {\bibfnamefont {H.}~\bibnamefont
  {Cram{\'e}r}},\ }\href@noop {} {\emph {\bibinfo {title} {Mathematical Methods
  of Statistics (PMS-9)}}},\ Vol.~\bibinfo {volume} {9}\ (\bibinfo  {publisher}
  {Princeton university press},\ \bibinfo {year} {2016})\BibitemShut {NoStop}%
\bibitem [{\citenamefont {Rao}(1992)}]{rao1992information}%
  \BibitemOpen
  \bibfield  {author} {\bibinfo {author} {\bibfnamefont {C.~R.}\ \bibnamefont
  {Rao}},\ }\bibfield  {title} {\enquote {\bibinfo {title} {Information and the
  accuracy attainable in the estimation of statistical parameters},}\ }in\
  \href {https://doi.org/10.1007/978-1-4612-0919-5_16} {\emph {\bibinfo
  {booktitle} {Breakthroughs in Statistics}}}\ (\bibinfo  {publisher}
  {Springer},\ \bibinfo {year} {1992})\ pp.\ \bibinfo {pages}
  {235--247}\BibitemShut {NoStop}%
\bibitem [{\citenamefont {Paris}(2009)}]{paris2009quantum}%
  \BibitemOpen
  \bibfield  {author} {\bibinfo {author} {\bibfnamefont {M.~G.~A.}\
  \bibnamefont {Paris}},\ }\bibfield  {title} {\enquote {\bibinfo {title}
  {Quantum estimation for quantum technology},}\ }\href
  {https://doi.org/10.1142/S0219749909004839} {\bibfield  {journal} {\bibinfo
  {journal} {Int. J. Quantum Inform.}\ }\textbf {\bibinfo {volume} {7}},\
  \bibinfo {pages} {125--137} (\bibinfo {year} {2009})}\BibitemShut {NoStop}%
\bibitem [{\citenamefont {Helstrom}(1969)}]{Helstrom1969}%
  \BibitemOpen
  \bibfield  {author} {\bibinfo {author} {\bibfnamefont {C.~W.}\ \bibnamefont
  {Helstrom}},\ }\bibfield  {title} {\enquote {\bibinfo {title} {Quantum
  detection and estimation theory},}\ }\href
  {http://dx.doi.org/10.1007/BF01007479} {\bibfield  {journal} {\bibinfo
  {journal} {J. Stat. Phys.}\ }\textbf {\bibinfo {volume} {1}},\ \bibinfo
  {pages} {231--252} (\bibinfo {year} {1969})}\BibitemShut {NoStop}%
\bibitem [{\citenamefont {Braunstein}\ and\ \citenamefont
  {Caves}(1994)}]{PhysRevLett.72.3439}%
  \BibitemOpen
  \bibfield  {author} {\bibinfo {author} {\bibfnamefont {S.~L.}\ \bibnamefont
  {Braunstein}}\ and\ \bibinfo {author} {\bibfnamefont {C.~M.}\ \bibnamefont
  {Caves}},\ }\bibfield  {title} {\enquote {\bibinfo {title} {Statistical
  distance and the geometry of quantum states},}\ }\href
  {http://dx.doi.org/10.1103/PhysRevLett.72.3439} {\bibfield  {journal}
  {\bibinfo  {journal} {Phys. Rev. Lett.}\ }\textbf {\bibinfo {volume} {72}},\
  \bibinfo {pages} {3439--3443} (\bibinfo {year} {1994})}\BibitemShut {NoStop}%
\bibitem [{\citenamefont {Braunstein}\ \emph {et~al.}(1996)\citenamefont
  {Braunstein}, \citenamefont {Caves},\ and\ \citenamefont
  {Milburn}}]{braunstein1996generalized}%
  \BibitemOpen
  \bibfield  {author} {\bibinfo {author} {\bibfnamefont {S.~L.}\ \bibnamefont
  {Braunstein}}, \bibinfo {author} {\bibfnamefont {C.~M.}\ \bibnamefont
  {Caves}}, \ and\ \bibinfo {author} {\bibfnamefont {G.~J.}\ \bibnamefont
  {Milburn}},\ }\bibfield  {title} {\enquote {\bibinfo {title} {Generalized
  uncertainty relations: theory, examples, and lorentz invariance},}\ }\href
  {https://doi.org/10.1006/aphy.1996.0040} {\bibfield  {journal} {\bibinfo
  {journal} {Ann. Phys.}\ }\textbf {\bibinfo {volume} {247}},\ \bibinfo {pages}
  {135--173} (\bibinfo {year} {1996})}\BibitemShut {NoStop}%
\bibitem [{\citenamefont {Helstrom}(1967)}]{HELSTROM1967101}%
  \BibitemOpen
  \bibfield  {author} {\bibinfo {author} {\bibfnamefont {C.~W.}\ \bibnamefont
  {Helstrom}},\ }\bibfield  {title} {\enquote {\bibinfo {title} {Minimum
  mean-squared error of estimates in quantum statistics},}\ }\href
  {http://dx.doi.org/10.1016/0375-9601(67)90366-0} {\bibfield  {journal}
  {\bibinfo  {journal} {Phys. Lett. A}\ }\textbf {\bibinfo {volume} {25}},\
  \bibinfo {pages} {101 -- 102} (\bibinfo {year} {1967})}\BibitemShut {NoStop}%
\bibitem [{\citenamefont {Helstrom}\ and\ \citenamefont
  {Kennedy}(1974)}]{helstrom1974noncommuting}%
  \BibitemOpen
  \bibfield  {author} {\bibinfo {author} {\bibfnamefont {C.~W.}\ \bibnamefont
  {Helstrom}}\ and\ \bibinfo {author} {\bibfnamefont {R.}~\bibnamefont
  {Kennedy}},\ }\bibfield  {title} {\enquote {\bibinfo {title} {Noncommuting
  observables in quantum detection and estimation theory},}\ }\href
  {https://doi.org/10.1109/TIT.1974.1055173} {\bibfield  {journal} {\bibinfo
  {journal} {IEEE Trans. Inf. Theory}\ }\textbf {\bibinfo {volume} {20}},\
  \bibinfo {pages} {16--24} (\bibinfo {year} {1974})}\BibitemShut {NoStop}%
\bibitem [{\citenamefont {Yuen}\ and\ \citenamefont {Lax}(1973)}]{Yuen1973}%
  \BibitemOpen
  \bibfield  {author} {\bibinfo {author} {\bibfnamefont {H.}~\bibnamefont
  {Yuen}}\ and\ \bibinfo {author} {\bibfnamefont {M.}~\bibnamefont {Lax}},\
  }\bibfield  {title} {\enquote {\bibinfo {title} {Multiple-parameter quantum
  estimation and measurement of nonselfadjoint observables},}\ }\href
  {http://dx.doi.org/10.1109/TIT.1973.1055103} {\bibfield  {journal} {\bibinfo
  {journal} {IEEE Trans. Inf. Theory}\ }\textbf {\bibinfo {volume} {19}},\
  \bibinfo {pages} {740--750} (\bibinfo {year} {1973})}\BibitemShut {NoStop}%
\bibitem [{\citenamefont {Belavkin}(1976)}]{belavkin1976generalized}%
  \BibitemOpen
  \bibfield  {author} {\bibinfo {author} {\bibfnamefont {V.~P.}\ \bibnamefont
  {Belavkin}},\ }\bibfield  {title} {\enquote {\bibinfo {title} {Generalized
  uncertainty relations and efficient measurements in quantum systems},}\
  }\href {https://doi.org/10.1007/BF01032091} {\bibfield  {journal} {\bibinfo
  {journal} {Theor. Math. Phys.}\ }\textbf {\bibinfo {volume} {26}},\ \bibinfo
  {pages} {213--222} (\bibinfo {year} {1976})}\BibitemShut {NoStop}%
\bibitem [{\citenamefont {Petz}\ and\ \citenamefont
  {Ghinea}(2011)}]{petz2011introduction}%
  \BibitemOpen
  \bibfield  {author} {\bibinfo {author} {\bibfnamefont {D.}~\bibnamefont
  {Petz}}\ and\ \bibinfo {author} {\bibfnamefont {C.}~\bibnamefont {Ghinea}},\
  }\bibfield  {title} {\enquote {\bibinfo {title} {Introduction to quantum
  fisher information},}\ }in\ \href
  {https://doi.org/10.1142/9789814338745_0015} {\emph {\bibinfo {booktitle}
  {Quantum probability and related topics}}},\ Vol.~\bibinfo {volume} {1},\
  \bibinfo {editor} {edited by\ \bibinfo {editor} {\bibfnamefont
  {R.}~\bibnamefont {Rebolledo}}\ and\ \bibinfo {editor} {\bibfnamefont
  {M.}~\bibnamefont {Orszag}}}\ (\bibinfo  {publisher} {World Scientific},\
  \bibinfo {year} {2011})\ pp.\ \bibinfo {pages} {261--281}\BibitemShut
  {NoStop}%
\bibitem [{\citenamefont {Fujiwara}(1994{\natexlab{a}})}]{fujiwara1994linear}%
  \BibitemOpen
  \bibfield  {author} {\bibinfo {author} {\bibfnamefont {A.}~\bibnamefont
  {Fujiwara}},\ }\bibfield  {title} {\enquote {\bibinfo {title} {Linear random
  measurements of two non-commuting observables},}\ }\href
  {http://www.keisu.t.u-tokyo.ac.jp/research/techrep/data/1994/METR94-10.pdf}
  {\bibfield  {journal} {\bibinfo  {journal} {Math. Eng. Tech. Rep}\ }\textbf
  {\bibinfo {volume} {94}} (\bibinfo {year} {1994}{\natexlab{a}})}\BibitemShut
  {NoStop}%
\bibitem [{\citenamefont {Fujiwara}\ and\ \citenamefont
  {Nagaoka}(1999)}]{fujiwara1999estimation}%
  \BibitemOpen
  \bibfield  {author} {\bibinfo {author} {\bibfnamefont {A.}~\bibnamefont
  {Fujiwara}}\ and\ \bibinfo {author} {\bibfnamefont {H.}~\bibnamefont
  {Nagaoka}},\ }\bibfield  {title} {\enquote {\bibinfo {title} {An estimation
  theoretical characterization of coherent states},}\ }\href
  {http://dx.doi.org/10.1063/1.532962} {\bibfield  {journal} {\bibinfo
  {journal} {J. Math. Phys.}\ }\textbf {\bibinfo {volume} {40}},\ \bibinfo
  {pages} {4227--4239} (\bibinfo {year} {1999})}\BibitemShut {NoStop}%
\bibitem [{\citenamefont {Fujiwara}(1994{\natexlab{b}})}]{fujiwara1994multi}%
  \BibitemOpen
  \bibfield  {author} {\bibinfo {author} {\bibfnamefont {A.}~\bibnamefont
  {Fujiwara}},\ }\bibfield  {title} {\enquote {\bibinfo {title}
  {Multi-parameter pure state estimation based on the right logarithmic
  derivative},}\ }\href
  {http://www.keisu.t.u-tokyo.ac.jp/research/techrep/data/1994/METR94-09.pdf}
  {\bibfield  {journal} {\bibinfo  {journal} {Math. Eng. Tech. Rep}\ }\textbf
  {\bibinfo {volume} {94}},\ \bibinfo {pages} {94--10} (\bibinfo {year}
  {1994}{\natexlab{b}})}\BibitemShut {NoStop}%
\bibitem [{\citenamefont {Nagaoka}(2005)}]{nagaoka2005new}%
  \BibitemOpen
  \bibfield  {author} {\bibinfo {author} {\bibfnamefont {H.}~\bibnamefont
  {Nagaoka}},\ }\bibfield  {title} {\enquote {\bibinfo {title} {A new approach
  to cram{\'e}r-rao bounds for quantum state estimation},}\ }in\ \href
  {http://dx.doi.org/10.1142/9789812563071_0009} {\emph {\bibinfo {booktitle}
  {Asymptotic Theory Of Quantum Statistical Inference: Selected Papers}}}\
  (\bibinfo {year} {2005})\ pp.\ \bibinfo {pages} {100--112}\BibitemShut
  {NoStop}%
\bibitem [{\citenamefont {Gao}\ and\ \citenamefont {Lee}(2014)}]{Gao2014}%
  \BibitemOpen
  \bibfield  {author} {\bibinfo {author} {\bibfnamefont {Y.}~\bibnamefont
  {Gao}}\ and\ \bibinfo {author} {\bibfnamefont {H.}~\bibnamefont {Lee}},\
  }\bibfield  {title} {\enquote {\bibinfo {title} {Bounds on quantum
  multiple-parameter estimation with gaussian state},}\ }\href
  {http://dx.doi.org/10.1140/epjd/e2014-50560-1} {\bibfield  {journal}
  {\bibinfo  {journal} {Eur. Phys. J. D}\ }\textbf {\bibinfo {volume} {68}},\
  \bibinfo {pages} {347} (\bibinfo {year} {2014})}\BibitemShut {NoStop}%
\bibitem [{\citenamefont {Duan}\ \emph {et~al.}(2000)\citenamefont {Duan},
  \citenamefont {Giedke}, \citenamefont {Cirac},\ and\ \citenamefont
  {Zoller}}]{duan2000inseparability}%
  \BibitemOpen
  \bibfield  {author} {\bibinfo {author} {\bibfnamefont {L.-M.}\ \bibnamefont
  {Duan}}, \bibinfo {author} {\bibfnamefont {G.}~\bibnamefont {Giedke}},
  \bibinfo {author} {\bibfnamefont {J.~I.}\ \bibnamefont {Cirac}}, \ and\
  \bibinfo {author} {\bibfnamefont {P.}~\bibnamefont {Zoller}},\ }\bibfield
  {title} {\enquote {\bibinfo {title} {Inseparability criterion for continuous
  variable systems},}\ }\href {https://doi.org/10.1103/PhysRevLett.84.2722}
  {\bibfield  {journal} {\bibinfo  {journal} {Phys. Rev. Lett.}\ }\textbf
  {\bibinfo {volume} {84}},\ \bibinfo {pages} {2722} (\bibinfo {year}
  {2000})}\BibitemShut {NoStop}%
\bibitem [{\citenamefont {Gu}\ \emph {et~al.}(2012)\citenamefont {Gu},
  \citenamefont {Chrzanowski}, \citenamefont {Assad}, \citenamefont {Symul},
  \citenamefont {Modi}, \citenamefont {Ralph}, \citenamefont {Vedral},\ and\
  \citenamefont {Lam}}]{gu2012observing}%
  \BibitemOpen
  \bibfield  {author} {\bibinfo {author} {\bibfnamefont {M.}~\bibnamefont
  {Gu}}, \bibinfo {author} {\bibfnamefont {H.~M.}\ \bibnamefont {Chrzanowski}},
  \bibinfo {author} {\bibfnamefont {S.~M.}\ \bibnamefont {Assad}}, \bibinfo
  {author} {\bibfnamefont {T.}~\bibnamefont {Symul}}, \bibinfo {author}
  {\bibfnamefont {K.}~\bibnamefont {Modi}}, \bibinfo {author} {\bibfnamefont
  {T.~C.}\ \bibnamefont {Ralph}}, \bibinfo {author} {\bibfnamefont
  {V.}~\bibnamefont {Vedral}}, \ and\ \bibinfo {author} {\bibfnamefont {P.~K.}\
  \bibnamefont {Lam}},\ }\bibfield  {title} {\enquote {\bibinfo {title}
  {Observing the operational significance of discord consumption},}\ }\href
  {https://doi.org/10.1038/nphys2376} {\bibfield  {journal} {\bibinfo
  {journal} {Nature Physics}\ }\textbf {\bibinfo {volume} {8}},\ \bibinfo
  {pages} {671--675} (\bibinfo {year} {2012})}\BibitemShut {NoStop}%
\end{thebibliography}%

\appendix
\onecolumngrid

\section{Conversion of problem to semi-definite program (SDP)}
\label{sec_app_sdp}

We show that the problem of computing Holevo's bound for mean value
estimation of Gaussian states is a semi-definite program. We formulate
the original problem of finding $\Sigma_*$ into a dual form SDP.
Holevo's bound is the following:
\begin{program}{Holevo's bound}
\begin{align}
  \label{prog:original}
  \Sigma_* = &\min_{\mathbb{F} \in \mathbb{S}^4} \Tr{F^{-1}} \\
 \textnormal{subject to }\;  &0\le \mathbb{F}\leq \mathbb{C}\;,
\end{align}
\end{program}
where $\mathbb{S}^n$ is the set of $n \times n$ real symmetric matrices, $F= \mathbb{M} \mathbb{F} \mathbb{M}^\dagger$, and
$\mathbb{M}$ is a fixed real 2-by-4 matrix. Also $\mathbb{C}\coloneqq (1+\tfrac{1}{2}i\mathbb{D})^{-1}$ is a
fixed Hermitian 4-by-4 matrix. To cast this nonlinear optimisation
problem to an SDP, we use the standard trick of introducing an
auxiliary 2-by-2 real matrix $H$ that serves as an upper bound to
$F^{-1}$. So Holevo's bound becomes
\begin{program}
\begin{align}
  \Sigma_* = &\min_{\mathbb{F}\in \mathbb{S}^4,H\in \mathbb{S}^2} \Tr{H} \\
 \textnormal{subject to }\;  &0\le \mathbb{F}\leq \mathbb{C}\\
& H \geq F^{-1}\;.
\end{align}
\end{program}
Consider 
  \begin{align}
  \label{eq:14}
&  W(\mathbb{F},H) =
  \begin{bmatrix}
    H&I_2\\
    I_2&F
  \end{bmatrix} \geq 0\ \\
& \Leftrightarrow W/F = H-F^{-1} \geq 0\\
&\Leftrightarrow H \geq F^{-1},
  \end{align}
where $W/F$ is the Schur's complement of $F$ in $W$, and $I_n$ is the $n\times n$ identity matrix. We can
formulate the SDP for $\Sigma_*$ as:
\begin{align}
  \label{eq:2}
  \Sigma_* = \min_{\mathbb{F} \in \mathbb{S}^4,H \in \mathbb{S}^2 } {\Tr{H}}
\end{align}
subject to
\begin{align}
  \label{eq:13}
  \begin{bmatrix}
    H&I_2\\
    I_2& \mathbb{M} \mathbb{F} \mathbb{M}^\dagger
  \end{bmatrix} \oplus \mathbb{F}
    \oplus -\mathbb{F}
& \geq  0_4\oplus 0_4 \oplus -\mathbb{C} \\
\Leftrightarrow \underbrace{  \begin{bmatrix}
    H&0_2\\
    0_2& \mathbb{M} \mathbb{F} \mathbb{M}^\dagger
  \end{bmatrix} \oplus \mathbb{F}
    \oplus -\mathbb{F}}_{ \sum_j y_j B_j }
&\geq\underbrace{ 
\begin{bmatrix}
0_2&-I_2\\
-I_2&0_2
\end{bmatrix}
\oplus 0_4 \oplus -\mathbb{C} }_{C}\;,
\end{align}
where $0_n$ is the $n\times n$ zero matrix. We can decompose the LHS
into a sum $\sum_j y_j B_j$ where
$y=\begin{bmatrix}y_1&\ldots&y_{13}\end{bmatrix}^\intercal$ is a
vector of real numbers and $B_j$ are the 13 matrices given by:
\begin{align}
  \label{eq:15}
  B_j&=
  \begin{bmatrix}
    \mathbb{B}_j&0_2\\
    0_2& \mathbb{M} \mathbb{A}_j \mathbb{M}^\dagger
\end{bmatrix} \oplus
    \mathbb{A}_j \oplus -\mathbb{A}_j \text{ for } j=1,\ldots,
        13\;.
\end{align}
$\{\mathbb{A}_j\}$ are 10 real symmetric matrices that forms a basis for the set of $4\times4$ real symmetric matrices. Similarly, $\{\mathbb{B}_j\}$ are three real symmetric matrices that forms a basis for the set of $2\times2$ real symmetric matrices. They are given by the following:
\begin{alignat*}{5}
  \mathbb{A}_1 &=
  \begin{bmatrix}
    1&0&0&0\\
    0&0&0&0\\
    0&0&0&0\\
    0&0&0&0
  \end{bmatrix}& \quad
  \mathbb{A}_2 &=
  \begin{bmatrix}
    0&0&0&0\\
    0&1&0&0\\
    0&0&0&0\\
    0&0&0&0
  \end{bmatrix}& \quad
  \mathbb{A}_3 &=
  \begin{bmatrix}
    0&0&0&0\\
    0&0&0&0\\
    0&0&1&0\\
    0&0&0&0
  \end{bmatrix}& \quad
  \mathbb{A}_4 &=
  \begin{bmatrix}
    0&0&0&0\\
    0&0&0&0\\
    0&0&0&0\\
    0&0&0&1
  \end{bmatrix}&\quad
  \mathbb{A}_5 &=
  \begin{bmatrix}
    0&1&0&0\\
    1&0&0&0\\
    0&0&0&0\\
    0&0&0&0
  \end{bmatrix}\\
  \mathbb{A}_6 &=
  \begin{bmatrix}
    0&0&1&0\\
    0&0&0&0\\
    1&0&0&0\\
    0&0&0&0
  \end{bmatrix}&
  \mathbb{A}_7 &=
  \begin{bmatrix}
    0&0&0&1\\
    0&0&0&0\\
    0&0&0&0\\
    1&0&0&0
  \end{bmatrix}&
  \mathbb{A}_8 &=
  \begin{bmatrix}
    0&0&0&0\\
    0&0&1&0\\
    0&1&0&0\\
    0&0&0&0
  \end{bmatrix}&
  \mathbb{A}_9 &=
  \begin{bmatrix}
    0&0&0&0\\
    0&0&0&1\\
    0&0&0&0\\
    0&1&0&0
  \end{bmatrix}&
  \mathbb{A}_{10} &=
  \begin{bmatrix}
    0&0&0&0\\
    0&0&0&0\\
    0&0&0&1\\
    0&0&1&0
  \end{bmatrix}
\end{alignat*}
and $\mathbb{A}_j =0$ for $j=11,12,13$;
\begin{align}
  \label{eq:17}
    \mathbb{B}_{11} = 
  \begin{bmatrix}
    1&0\\
    0&0
  \end{bmatrix} \quad
    \mathbb{B}_{12} = 
  \begin{bmatrix}
    0&0\\
    0&1
  \end{bmatrix} \quad
    \mathbb{B}_{13} = 
  \begin{bmatrix}
    0&1\\
    1&0
  \end{bmatrix}\quad
       \mathbb{B}_j =0\; \textrm{ for } j=1,\ldots,10\;.
\end{align}
The objective function can be written as $\tr{H}=y^\intercal \,b$
where  $b=\begin{bmatrix}0&0&0&0&0&0&0&0&0&0&1&1&0 \end{bmatrix}^\intercal$.
Finally, we have the problem statement as the following:
\begin{program}{Standard SDP dual problem formulation of Holevo's
    bound}
\label{sdp_dual}
\begin{align}
\Sigma_*&= \min_y y^\intercal \,b\\
 \textnormal{subject to }\; &  \sum_j y_j B_j \geq C\;. \label{con_dual}
\end{align}
\end{program}
This is traditionally
called the dual problem.

The primal problem statement is:
\begin{program}{Standard SDP primal problem formulation of Holevo's bound}
\label{sdp_pri}
\begin{align}
  \Sigma^* &= \max_{X} \Tr{C X}\\
 \textnormal{subject to }\; &  \Tr{B_j X} = b_j \textnormal{ for }
                              j=1,\ldots,13, \label{con_pri}
\end{align}
\end{program}
where $X$ is a positive Hermitian matrix. This problem is bounded above and strictly feasible, which means that
it satisfies strong duality: $\Sigma_*=\Sigma^*$.

\section{Solution to the worked example}
\label{sec_app1}
In this appendix we provide the solution to the worked example. We
present $X^*$ and $y^*$ that we claim is optimal. We first verify that
$X^*$ and $y^*$ satisfy the primal and dual constraint. Next we show
that the primal and dual value they provide are the same, indicating
that the solution is optimal.

We consider the solutions for $r\ge r_0$ and $r<r_0$ separately.

\subsection{Solution for $r < r_0$}
For $r<r_0$, we claim that a solution
 is achieved by $y^*$ and $X^*$ having the form
\begin{align}
  \label{eq:1}
  y^*&=\begin{bmatrix}c_1&c_2&c_2&c_1&0&-c_0&0&0&-c_0&0&\frac{4v^2-1}{4v
      \cosh 2r-2}&\frac{4v^2-1}{4v \cosh
      2r-2}&0\end{bmatrix}^\intercal, \\
X^*&=X_1^* \oplus 0_4 \oplus X_3^*,
\end{align}
where we are free to choose $c_0:$ $0 \leq c_0 \leq
\frac{2v}{1-4v^2}+\frac{v}{2v\cosh2r-1}$ and
\begin{align}
  \label{eq:3}
  c_1&=\frac{2v(2v-\e^{-2r})}{4v^2-1}-\e^{-2r}c_0,\\
  c_2&=\frac{2v(2v-\e^{ 2r})}{4v^2-1}-\e^{ 2r}c_0,\\
  X_1^*&=
  \begin{bmatrix}
    1&0&-\frac{4v^2-1}{4v \cosh 2r-2}&0\\
    0&1&0&-\frac{4v^2-1}{4v \cosh 2r-2}\\
    -\frac{4v^2-1}{4v \cosh 2r-2}&0&\frac{(4v^2-1)^2}{(4v \cosh 2r-2)^2}&0\\
    0&-\frac{4v^2-1}{4v \cosh 2r-2}&0&\frac{(4v^2-1)^2}{(4v \cosh 2r-2)^2}
  \end{bmatrix}, \\
  X_3^*&=\frac{(4v^2-1)^2}{2v(4v \cosh 2r-2)^2}
  \begin{bmatrix}
    e^{2r}& i&-1&-i\, e^{2r}\\
    - i&e^{-2r}& i\,e^{-2r}&-1\\
    -1&- i\,e^{-2r}&e^{-2r}& i\\
    i\,e^{2r} &-1&- i&e^{2r}
  \end{bmatrix}\;.
\end{align}
Simple algebra confirms that $X^*$ satisfies $\Tr{B_j X^*}=b_j$, and
the nonzero eigenvalues of $X^*$ are
\begin{equation}
\left( \frac{(4v^2-1)^2\cosh 2r}{2v(2v \cosh 2r-1)^2},
1+ \frac{(4v^2-1)^2}{4(2v \cosh 2r-1)^2} \text{ (deg 2) }
\right),
\end{equation}
where (deg 2) indicates that the eigenvalue has degeneracy 2. The eigenvalues are nonnegative, so $X^*$ is a valid solution to the primal problem.

Now let us verify that $y^*$ satisfies the dual problem constraint~(\ref{con_dual}):
\begin{align}
\sum_j y^*_j B_j &= \begin{bmatrix}
\frac{4v^2-1}{4v\cosh 2r-2} & 0 &0&0\\
0&\frac{4v^2-1}{4v\cosh 2r-2}&0&0\\
0&0&\frac{4v\cosh 2r-2}{4v^2-1}&0\\
0&0&0&\frac{4v\cosh 2r-2}{4v^2-1}
\end{bmatrix} \oplus \begin{bmatrix}
c_1&0&-c_0&0\\
0&c_2&0&-c_0\\
-c_0&0&c_2&0\\
0&-c_0&0&c_1
\end{bmatrix} \oplus \begin{bmatrix}
-c_1&0&c_0&0\\
0&-c_2&0&c_0\\
c_0&0&-c_2&0\\
0&c_0&0&-c_1
\end{bmatrix}
\end{align}
where 
\begin{align}
\label{con_C}
\mathbb C &= \begin{bmatrix}
0&0&-1&0\\
0&0&0&-1\\
-1&0&0&0\\
0&-1&0&0
\end{bmatrix} \oplus
0_4
\oplus 
\frac{2v}{4v^2-1}
\begin{bmatrix}
-2v&i&0&0\\
-i&-2v&0&0\\
0&0&-2v&i\\
0&0&-i&-2v
\end{bmatrix}\;.
\end{align}
The nonzero eigenvalues of $\sum_j
y_j^* \mathbb B_j-\mathbb C$ are then
\begin{align}
  \begin{bmatrix}
    \frac{4v^2-1}{4v\cosh 2r-2}+\frac{4v \cosh 2r-2}{4v^2-1}
    \;\textrm{ (deg 2) }\\
2 \left(c_0 + \frac{2v}{4v^2-1} \right)  \cosh 2r\\
\frac{\cosh 2r}{4v^2-1} \left( 2v+ (4v^2-1)c_0 + \sqrt{\left(
2v+(4v^2-1)c_0  \right)^2-\frac{8 v(4v^2-1)c_0}{\cosh^2 2r}}\right)\\
\frac{\cosh 2r}{4v^2-1} \left( 2v+ (4v^2-1)c_0 - \sqrt{\left(
2v+(4v^2-1)c_0  \right)^2-\frac{8 v(4v^2-1)c_0}{\cosh^2 2r}}\right)  \\
\frac{1}{4v^2-1}\left(4v^2 -\left(2v  + (4v^2-1)c_0\right) \cosh 2r+
\sqrt{\left( 2v + (4v^2-1) c_0\right)^2 \cosh^2 2r-
    4v^2-4v (4v^2-1)c_0} \right)   \;\textrm{ (deg 2) }\\
\frac{1}{4v^2-1}\left(4v^2 -\left(2v  + (4v^2-1)c_0\right) \cosh 2r-
\sqrt{\left( 2v + (4v^2-1) c_0\right)^2 \cosh^2 2r-
    4v^2-4v (4v^2-1)c_0} \right)   \;\textrm{ (deg 2) }
  \end{bmatrix}
\end{align}
The first five eigenvalues are positive when $v\geq
\frac{1}{2}$ and $c_0\geq 0$, while the last is positive when $c_0\leq \frac{2v}{1-4v^2}+\frac{v}{2v\cosh2r-1}$.

The value of the dual is
$y^\intercal b=\frac{4v^2-1}{2v \cosh 2r-1}$. It can be verified
using simple algebra that the primal value $\Tr{C X^*}$ is also equal to
$\frac{4v^2-1}{2v \cosh 2r-1}$. Since the primal is equal to the
dual, we know that the solution is optimal.

One might wonder why the optimal measurement does not depend on
$c_0$. Any $c_0$ would give rise to an $\mathbb{F}_*$ that is optimal,
\begin{align}
  \mathbb{F}_* = 
 \begin{bmatrix}
c_1&0&-c_0&0\\
0&c_2&0&-c_0\\
-c_0&0&c_2&0\\
0&-c_0&0&c_1
\end{bmatrix} ,
\end{align}
 and hence different $\mathscr{F}_*$; however, the
vectors $\mathscr{F}_* m_j$ does not depend on $c_0$. By direct
computation
\begin{align}
  \mathscr{F}_* m_1 = \frac{\sqrt{2}}{4v^2-1}
  \begin{bmatrix}
    2v\e^{2r}-1\\0\\1-2v \e^{-2r}\\0
  \end{bmatrix}, \\
  \mathscr{F}_* m_2 = \frac{\sqrt{2}}{4v^2-1}
  \begin{bmatrix}
    2v\e^{-2r}-1\\0\\1-2v \e^{2r}\\0
  \end{bmatrix}
\end{align}
is independent of $c_0$.

\subsection{Solution for $r\ge r_0$}
When $r \ge r_0$, we claim that the optimal values of $X$ and $y$ that attains $h_*$ and $g_*$ in the SDP program (\ref{sdp_dual}) and
(\ref{sdp_pri}) are given by
\begin{align}
  \label{eq:1}
  y^*&=\begin{bmatrix}1&0&0&1&0&0&0&0&0&0&2v \e^{-2r}&2v \e^{-2r}&0\end{bmatrix}^\intercal, \\
X^*&=X_1^* \oplus X_2^* \oplus X_3^*,
\end{align}
where
\begin{align}
  \label{eq:3}
  X_1^*&=
  \begin{bmatrix}
    1&0&-2v\e^{-2r}&0\\
    0&1&0&-2v\e^{-2r}\\
    -2v\e^{-2r}&0&4v^2\e^{-4r}&0\\
    0&-2v\e^{-2r}&0&4v^2\e^{-4r}
  \end{bmatrix}, \\
  X_2^*&=\frac{e^{-2r}(1-4v^2 \e^{-4r})}{2v}
  \begin{bmatrix}
    0&0&0&0\\
    0&1&0&0\\
    0&0&1&0\\
    0&0&0&0
  \end{bmatrix}, \\
  X_3^*&=\e^{-2r}
  \begin{bmatrix}
    2v& i&-2v\e^{-2r}&- 4v^2 \e^{-2r}i\\
    - i&\frac{1}{2v}&\e^{-2r} i&-2v\e^{-2r}\\
    -2v\e^{-2r}&-\e^{-2r} i&\frac{1}{2v}& i\\
       4v^2 \e^{-2r} i&-2v\e^{-2r}&- i&2v
  \end{bmatrix}.
\end{align}
To justify this claim, we need to show that $X^*$ and $y^*$ satisfies
constraints (\ref{con_dual}) and (\ref{con_pri}) and that the value of the dual
solution is equal to the primal solution, $\Sigma_*=\Sigma^*$.

To check the constraint for the dual (\ref{con_dual}), we compute the eigenvalues of $\sum_j y^*_j
B_j-\mathbb C$ where
\begin{align}
\sum_j y^*_j B_j &= \begin{bmatrix}
2v\e^{-2r} & 0 &0&0\\
0&2v\e^{-2r}&0&0\\
0&0&\frac{\e^{2r}}{2v}&0\\
0&0&0&\frac{\e^{2r}}{2v}
\end{bmatrix} \oplus \begin{bmatrix}
1&0&0&0\\
0&0&0&0\\
0&0&0&0\\
0&0&0&1
\end{bmatrix} \oplus \begin{bmatrix}
-1&0&0&0\\
0&0&0&0\\
0&0&0&0\\
0&0&0&-1
\end{bmatrix},
\end{align}
and $\mathbb{C}$ is given in~(\ref{con_C}). The nonzero eigenvalues of  $\sum_j y^*_j
B_j-C$ are
\begin{equation}
\left(1,\frac{1+4v^2 \e^{-4r}}{2v\e^{-2r}},\frac{4v^2+1}{4v^2-1}
\right)\;,
\end{equation}
each occurring with degeneracy two. Since $v > \frac{1}{2}$, so all of
the eigenvalues are nonnegative. Hence $y^*$ is a valid solution.

Simple algebra confirms that the primal constraint $\Tr{B_j X^*}=b_j$
is also satisfied. The nonzero eigenvalues of $X^*$ are 
\begin{equation}
\left(
1+4v^2\e^{-4r} \textrm{ (deg 2) },
\frac{(1-4v^2 \e^{-4r})\e^{-2r}}{2v} \textrm{ (deg 2) },
\frac{(1+4v^2)(1-2v \e^{-2r})\e^{-2r}}{2v} ,
\frac{(1+4v^2)(1+2v \e^{-2r})\e^{-2r}}{2v} 
 \right)
\end{equation}
All of these eigenvalues are nonnegative
provided $\e^{2r}-2v\ge 0$, which is just the condition for $r \ge
r_0$. Therefore $X^*$ is positive definite when $r \ge r_0$, and the
constraints for the primal problem are satisfied. Therefore we have
shown that $y^*$ and $X^*$ specified above are a valid solution.

Next, by direct computation, $y_*^\intercal b =4v\e^{-2r}$ and also
$\Tr{C X^*} =4v \e^{-2r}$. Since the primal is equal to the
dual, the solution is optimal.

\section{Generalization to $n$-mode states}

To generalize the results in Sec.~\ref{sec_calculation} to an $n$-mode Gaussian
state, we extend the definition of $z$ to
$z=\begin{bmatrix}y_1&x_1&y_2&x_2&...&y_n&x_n\end{bmatrix}^\intercal$
in a $2n$-dimensional real vector space $Z$ and the canonical observables
\begin{align}
  \mathcal{R}(z) = \sum_j x_j \mathcal{P}_j + y_j \mathcal{Q}_j \;,
\end{align}
where $\mathcal{P}_j$ and $\mathcal{Q}_j$ are the quadrature operators
for the $j$-th mode. The skew-symmetric bilinear form generalises to
\begin{align}
 \Delta(z,z') = \sum_j x'_j y_j - x_j y'_j
\end{align}
such that the commutation relation Eq.\ (\ref{eq_commutation}) still holds. Equation (\ref{eq_mz})
defining the mean value function and Eq.\ (\ref{eq_alpha}) defining the correlation function of
the Gaussian state remains unchanged. To estimate $l$ displacement
parameters $\theta_j$ for $j=1, \ldots, l$, we introduce $m_j(z)$ for
$j=1,\ldots,l$ such that
\begin{align}
  m(z) = \sum_j \theta_j m_j(z).
\end{align}
The results of Sec.~\ref{sec_holevobound} then follow with only
minor modification to the size of the matrix $F$ which is now
$l$-by-$l$. The definitions and results of Secs.~\ref{sec_optimal_measure}, \ref{sec_matrix_rep}, and Appendix \ref{sec_app_sdp} are
still valid after appropriately extending the matrix and vector dimensions.

\end{document}